%% file: ms.tex
\documentclass[10pt,conference,letterpaper]{IEEEtran}

\usepackage{booktabs} 
\usepackage{amsmath}
\usepackage{subfigure}
\usepackage{times}
\let\labelindent\relax
\usepackage[shortlabels]{enumitem}
\usepackage[ruled, vlined, linesnumbered]{algorithm2e}
\usepackage{algorithmicx}
\usepackage{algpseudocode}
\usepackage{multirow}
\usepackage[pdftex]{graphicx}
\usepackage{url}
\usepackage{color}

\newtheorem{definition}{Definition}
\newtheorem{assumption}{Assumption}
\newtheorem{theorem}{Theorem}

\newcommand{\figcaptionspace}{0pt}

\title{Enabling Quality Control for Entity Resolution: A Human and Machine Cooperation Framework \vspace{-0.4cm}}

\author{
{
\resizebox{\linewidth}{!}{
\mbox{Zhaoqiang Chen{\small $~^{\#1}$}, Qun Chen{\small $~^{\#2}$}, Fengfeng Fan{\small $~^{\#3}$}, Yanyan Wang{\small $~^{\#4}$}, Zhuo Wang{\small $~^{\#5}$}, Youcef Nafa{\small $~^{\#6}$}, Zhanhuai Li{\small $~^{\#7}$}, Hailong Liu{\small $~^{\#8}$}, Wei Pan{\small $~^{\#9}$}
}}
}%
\vspace{1.6mm}\\
\fontsize{10}{10}\selectfont\itshape
$^{\#}$\,School of Computer Science, Northwestern Polytechnical University\\
$^{\#}$\,Key Laboratory of Big Data Storage and Management, Northwestern Polytechnical University, Ministry \\ of Industry and Information Technology\\
1 Dongxiang Road, Xi'an Shaanxi, China\\
\fontsize{9}{9}\selectfont\ttfamily\upshape
\{$^{2}$\,chenbenben, $^{7}$\,lizhh, $^{8}$\,liuhailong, $^{9}$\,panwei1002 \}@nwpu.edu.cn\\
\{$^{1}$\,chenzhaoqiang, $^{3}$\,fanfengfeng, $^{4}$\,wangyanyan, $^{5}$\,wzhuo918, $^{6}$\,youcef.nafa \}@mail.nwpu.edu.cn\\
}
\begin{document}

\maketitle
\begin{abstract}
Even though many machine algorithms have been proposed for entity resolution, it remains very challenging to find a solution with quality guarantees. In this paper, we propose a novel HUman and Machine cOoperation (HUMO) framework for entity resolution (ER), which divides an ER workload between the machine and the human. HUMO enables a mechanism for quality control that can flexibly enforce both precision and recall levels. We introduce the optimization problem of HUMO, minimizing human cost given a quality requirement, and then present three optimization approaches: a conservative baseline one purely based on the monotonicity assumption of precision, a more aggressive one based on sampling and a hybrid one that can take advantage of the strengths of both previous approaches. Finally, we demonstrate by extensive experiments on real and synthetic datasets that HUMO can achieve high-quality results with reasonable return on investment (ROI) in terms of human cost, and it performs considerably better than the state-of-the-art alternatives in quality control.
\end{abstract}

%
%





\input{1-introduction}
\input{5-relatedwork-chen}

\input{2-framework}

\input{3-baseline}

\input{3-sampling}
\input{3-hybrid}
\input{4-experiments}

\input{6-conclusion}
\input{7-acknowledgements}

\bibliographystyle{IEEEtran}
\bibliography{mybibfile}

\end{document}

%% file: 1-introduction.tex
\section{introduction}
  Entity resolution (ER) usually refers to identifying the relational records that correspond to the same real-world entity in a dataset. Extensively studied in the literature \cite{christen2012data}, ER can be performed based on rules \cite{fan2009reasoning, li2015rule, singh2017generating}, probabilistic theory \cite{fellegi1969theory} or machine learning \cite{sarawagi2002interactive, kouki2017collective, arasu2010active, bellare2012active}. Unfortunately, most of the existing techniques fall short of effective mechanisms for quality control. As a result, they cannot enforce quality guarantees. Even though active learning based approaches \cite{arasu2010active, bellare2012active} can optimize recall while ensuring a user-specified precision level, it is usually desirable in practice that an ER result can have more comprehensive quality guarantees specified at both precision and recall fronts.

  To flexibly impose quality guarantees, we propose a novel human and machine cooperation framework, HUMO, for ER. Its primary idea is to divide instance pairs in an ER workload into easy ones, which can be automatically labeled by a machine with high accuracy; and more challenging ones, which require manual verification. HUMO is, to some extent, inspired by the success of human intervention in problem solving as demonstrated by numerous crowdsourcing applications \cite{li2016crowdsourced}. However, existing crowdsourcing solutions for ER \cite{wang2012crowder, whang2013question, vesdapunt2014crowdsourcing, gokhale2014corleone, mozafari2014scaling, wang2015crowd, chai2016cost} mainly focused on how to make humans more effective and efficient on a given workload. Targeting the challenge of quality control, HUMO instead investigates the problem of how to divide an ER workload between the human and the machine such that a given quality requirement can be met.

  HUMO is motivated by the observation that pure machine-based solutions usually struggle in ensuring desired quality guarantees for tasks as challenging as entity resolution. Even though humans usually perform better than machines in terms of quality on such tasks, human labor is much more expensive than machine computation. Therefore, HUMO has been designed with the purpose of minimizing human cost given a particular quality requirement. Note that a prototype system of HUMO has been demonstrated in~\cite{chen2017humo}. The major contributions of this technical paper can be summarized as follows:

\begin{enumerate}
\item We propose a human and machine cooperation framework, HUMO, for entity resolution. The attractive property of HUMO is that it enables an effective mechanism for comprehensive quality control at both precision and recall fronts;
\item We introduce the optimization problem of HUMO, i.e. minimizing human cost given a quality requirement, and present three optimization approaches: a conservative baseline one purely based on the monotonicity assumption of precision, a more aggressive one based on sampling, and a hybrid one that can take advantage of the strengths of both approaches;
\item We validate the efficacy of HUMO by extensive experiments on both real and synthetic datasets. Our empirical evaluation shows that HUMO can achieve high-quality results with reasonable ROI in terms of human cost, and it performs considerably better than the state-of-the-art alternatives in quality control. On minimizing human cost, the hybrid approach performs better than both the baseline and sampling-based approaches.
\end{enumerate}

  The rest of this paper is organized as follows: Section~\ref{sec:related} reviews more related work. Section~\ref{sec:setting} defines the task.  Section~\ref{sec:framework} presents the framework. Section~\ref{sec:conservative} describes the baseline approach based on the monotonicity assumption of precision. Section~\ref{sec:aggressive} describes the sampling-based approach. Section~\ref{sec:hybrid} describes the hybrid approach. Section~\ref{sec:experiment} presents our empirical evaluation results.  Finally, Section~\ref{sec:conclusion} concludes this paper with some thoughts on future works.

%% file: 5-relatedwork-chen.tex
\section{related work} \label{sec:related}

  As a classical problem in the area of data quality, entity resolution has been extensively studied in the literature~\cite{christen2012data, elmagarmid2007duplicate, christophides2015entity}. The proposed techniques include those based on rules~\cite{fan2009reasoning, li2015rule, singh2017generating}, probabilistic theory~\cite{fellegi1969theory, singla2006entity} and machine learning~\cite{sarawagi2002interactive, kouki2017collective, arasu2010active, bellare2012active}. However, these traditional techniques lack effective mechanisms for quality control; ergo, they fail in ensuring high-quality guarantees.

  Active learning-based approaches~\cite{arasu2010active, bellare2012active} have been proposed in order to satisfy the precision requirement for ER. The authors of \cite{arasu2010active} proposed a technique that can optimize the recall while ensuring a pre-specified precision goal. The authors in~\cite{bellare2012active} proposed an improved algorithm that approximately maximizes the recall under a precision constraint. Considering that these techniques share the same classification paradigm with traditional machine learning-based ones; the former cannot enforce comprehensive quality guarantees specified by both precision and recall metrics as HUMO does.

   The progressive paradigm for ER~\cite{whang2013pay, altowim2014progressive} has also been proposed for the application scenarios in which ER should be processed efficiently, but it does not necessarily guarantee high-quality results. Taking a pay-as-you-go approach, it studied how to maximize the result's quality given a pre-specified resolution budget, which was defined based on the machine computation cost. A similar iterative algorithm, SiGMa, was proposed in \cite{lacoste2013sigma}. It can leverage both the structure information and string similarity measures to resolve entity alignment across different knowledge bases. Note that built on machine computation, these techniques could not be applied to enforce quality guarantees either.

   It has been well recognized that pure machine algorithms may not be able to produce satisfactory results in many practical scenarios~\cite{li2016crowdsourced}. Many researchers~\cite{wang2012crowder, whang2013question, vesdapunt2014crowdsourcing, gokhale2014corleone, mozafari2014scaling, wang2015crowd, chai2016cost, verroios2017waldo} have studied how to crowdsource an ER workload. For instance, recently, the authors of~\cite{chai2016cost} proposed a cost-effective framework that employs the partial order relationship on instance pairs to reduce the number of asked pairs. Similarly, the authors in~\cite{verroios2017waldo} provided solutions to take advantage of both pairwise and multi-item interfaces in a crowdsourcing setting. While, these works addressed the challenges specific to crowdsourcing; we instead investigate a different problem: how to divide a workload between the human and the machine such that the user-specified quality guarantees can be met. In this paper, we assume that human workload can be performed with high quality; yet we do not investigate the problems targeted by existing interactive and crowdsourcing solutions. Note that the workload assigned to the human by HUMO can be naturally processed in a crowdsourcing manner. Our work can thus be considered orthogonal to existing works on crowdsourcing. It is interesting to investigate how to seamlessly integrate a crowdsourcing platform into HUMO in future work.

%% file: 2-framework.tex
\section{Problem Setting} \label{sec:setting}

  Entity resolution's main purpose is to determine whether two records are equivalent. Two records are deemed equivalent if and only if they correspond to the same real-world entity. We denote an ER workload by $D$, $D=\{d_1, d_2, \cdots, d_n\}$, in which $d_i$ represents an instance pair.  An ER solution corresponds to a label assignment $L$ for $D$, $L=\{l_1, l_2, \cdots, l_n\}$, in which $l_i=1$ if $d_i$ is labeled as {\em match} and $l_i=0$ if it is labeled as {\em unmatch}. In this paper, $d_i$ is called a matching pair if its two records are equivalent; otherwise, it is called an unmatching pair.

  As usual, we measure the quality of an ER solution by the metrics of precision and recall. Precision is the fraction of matching pairs among the pairs labeled as {\em match}, while recall is the fraction of correctly labeled matching pairs among all the matching pairs. Formally, we denote the ground-truth labeling solution for $D$ by $\hat{L}$, $\hat{L} = \{\hat{l}_1, \hat{l}_2, \cdots, \hat{l}_n\}$, in which $\hat{l}_i=1$ if $d_i$ is a matching pair and $\hat{l}_i=0$ otherwise. Given a labeling solution $L$, we use $D_{tp}$ to denote its set of true positive pairs, $D_{tp}=\{d_i|\hat{l}_i=1 \wedge l_i=1 \}$, $D_{fp}$ to denote its set of false positive pairs, $D_{fp}=\{d_i|\hat{l}_i=0 \wedge l_i=1 \}$, and $D_{fn}$ to denote its set of false negative pairs, $D_{fn}=\{d_i|\hat{l}_i=1 \wedge l_i=0\}$. Accordingly, the achieved precision level of $L$ can be represented by
\begin{equation} \label{eq:precision}
    \emph{precision(D,L)} = \frac{|D_{tp}|}{|D_{tp}| + |D_{fp}|}.
\end{equation}
Similarly, the achieved recall level of $L$ can be represented by
\begin{equation} \label{eq:recall}
    \emph{recall(D,L)} = \frac{|D_{tp}|}{|D_{tp}| + |D_{fn}|}.
\end{equation}

   Formally, the problem of entity resolution with quality guarantees specified at both precision and recall fronts is defined as follows:
\begin{definition}
\label{problemsetting}
{\bf [Entity Resolution with Quality Guarantees]}  Given a set of instance pairs, $D=\{d_1, d_2, \cdots, d_n\}$, the problem of entity resolution with quality guarantees is to give a labeling solution $L$ for $D$ provided with a confidence level $\theta$, $precision(D,L)\geq\alpha$ and $recall(D,L)\geq\beta$, in which $\alpha$ and $\beta$ denote the user-specified precision and recall values respectively.
\end{definition}

\section{HUMO Framework} \label{sec:framework}

  In this section, we first give an overview on HUMO, then introduce its optimization problem.

\subsection{Framework Overview}

    The primary idea behind HUMO is to enforce quality guarantees by dividing an ER workload between the human and the machine. It assigns easy instances, which can be automatically labeled with high accuracy, to the machine, while leaving more challenging instances for human-operated manual verification.

		Suppose that each instance pair in $D$ can be evaluated by a machine metric. This metric can be pair similarity or other classification metrics (e.g. match probability \cite{fellegi1969theory} and Support Vector Machine distance \cite{kopcke2010evaluation}). Note that entity resolution by classification usually categorizes pairs into $match$ and $unmatch$ based on a selected metric. Given a machine metric, HUMO assumes that $D$ statistically satisfies monotonicity of precision. Given a set of instance pairs, its precision refers to the proportion of matching pairs among all pairs. Intuitively, the monotonicity assumption of precision states that the higher (resp. lower) metric values a set of pairs have, the more probably they are matching pairs (resp. unmaching pairs). It can be observed that given a machine metric, the monotonicity assumption of precision underlies its effectiveness as a classification metric. {\em For simplicity of presentation, we use pair similarity as a machine metric example in this paper. However, HUMO is similarly effective with other machine metrics}. For instance, with the metric of SVM, each pair can be measured by its distance to a classification plane; with the metric of match probability, each pair can be measured by its estimated probability.

 Formally, we define the monotonicity assumption of precision, which was first proposed in \cite{arasu2010active}, as follows:
\begin{assumption}[Monotonicity of Precision]
  A value interval $I_i$ is dominated by another interval $I_j$, denoted by $I_i\preceq I_j$, if every value in $I_i$ is less than every value in $I_j$. We say that precision is monotonic with respect to a pair metric if for any two value intervals $I_i\preceq I_j$ in [0,1], we have $\mathsf{R}(I_i)\leq\mathsf{R}(I_j)$, in which $\mathsf{R}(I_i)$ denotes the precision of the set of instance pairs whose metric values are located in $I_i$.
\label{monotonicity}
\end{assumption}

\begin{figure}[h]
\setlength{\abovecaptionskip}{\figcaptionspace}
\centering
\includegraphics[width=\linewidth]{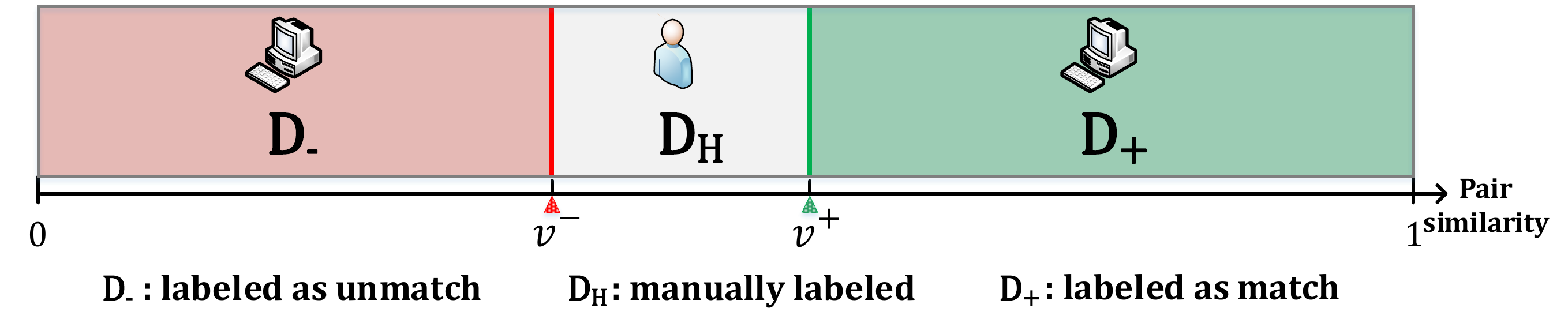}
\caption{The HUMO framework.}
\label{fig_basic_idea}
\end{figure}

  With the metric of pair similarity, the underlying intuition of Assumption \ref{monotonicity} is that the more similar two records are, the more likely they refer to the same real-world entity. According to the monotonicity assumption, a pair with high similarity has a correspondingly high probability of being a matching pair. A pair with low similarity instead has a correspondingly low probability of being a matching pair. These two groups of instance pairs can be supposed to be easy in that they can be automatically labeled by the machine with high accuracy. In comparison, the instance pairs having medium similarities are more challenging because labeling them either way by machine would introduce more considerable errors.

  The HUMO framework is shown in Figure~\ref{fig_basic_idea}. It divides the similarity interval [0,1] into three disjoint intervals, $I_-$, $I_H$ and $I_+$, in which $I_-$=[0,$v^-$), $I_H$=[$v^-$,$v^+$] and $I_+$=($v^+$,1], and correspondingly $D$ into three disjoint subsets, $D_-$, $D_H$ and $D_+$. It automatically labels the pairs in $D_-$ as {\em unmatch}, the pairs in $D_+$ as {\em match}, and assigns the pairs in $D_H$ to the human for manual verification. It can be observed that HUMO can flexibly enforce quality guarantees by adjusting the range of $D_H$. In the extreme case of $D_H=\emptyset$, HUMO boils down to a straightforward machine-based classification technique. With the assumption that the human performs better than the machine on a quality basis, enlarging the range of $D_H$ would result in improved quality. In the opposite extreme case of $D_H=D$, HUMO achieves the best performance, which is the same as the human's.

  Generally, given a HUMO solution $S$ consisting of $D_-$, $D_H$ and $D_+$, the lower bound of its achieved precision level can be represented by
\begin{equation}
   precision_l(S)=\frac{N^+_l(D_+)+N^+_l(D_H)}{N(D_+)+N(D_H)},
\label{eq:precision-bound}
\end{equation}
in which $N(\cdot)$ denotes the total number of pairs in a set and $N^+_l(\cdot)$ denotes the lower bound of the total number of matching pairs in a set. Similarly, the lower bound of its achieved recall level can be represented by
\begin{equation}
  recall_l(S)=\frac{N^+_l(D_+)+N^+_l(D_H)}{N^+_l(D_+)+N^+_l(D_H)+N^+_u(D_-)},
\label{eq:recall-bound}
\end{equation}
in which $N^+_u(\cdot)$ denotes the upper bound of the total number of matching pairs in a set. In this paper, for the sake of presentation simplicity, we assume that the pairs in $D_H$ can be manually labeled accurately (100\% accuracy with 100\% confidence). With that being said, we emphasize that HUMO's effectiveness does not depend on said assumption, since it can work properly provided that quality guarantees can be enforced on $D_H$. In the case that human errors are introduced in $D_H$, the lower bounds of the achieved precision and recall can be similarly estimated based on Eq.~\ref{eq:precision-bound} and Eq.~\ref{eq:recall-bound}. Nonetheless, it is worthy to point out that under the assumption that the human yields higher quality matches than the machine, the best quality guarantees HUMO can achieve are no better than human attained ones on $D_H$.

\subsection{Optimization Problem}

  Since human labor is in practice much more expensive than machine computation, HUMO aims to minimize human cost provided that user-specified quality requirements can be satisfied. By quantifying human cost by the number of manually inspected instance pairs in $D_H$, we formally define HUMO's optimization problem as follows:

\begin{definition}
\label{optimization}
{\bf [Minimizing Human Cost in HUMO].} Given a set of instance pairs, $D$, a confidence level $\theta$, a precision level $\alpha$ and a recall level $\beta$, HUMO's optimization problem is represented by
\begin{equation}
\begin{split}
& \quad \underset{S_i}{argmin} (|D_H(S_i)|)\\
& subject \quad to \quad P(precision(D,S_i)\geq\alpha)\geq\theta , \\
& \hspace{0.7in} P(recall(D,S_i)\geq\beta)\geq\theta ,
\end{split}
\label{eq:minimization}
\end{equation}
in which $S_i$ denotes a HUMO solution, $D_H(S_i)$ denotes the set of instance pairs assigned to the human by $S_i$, $precision(D,S_i)$ denotes the achieved precision by $S_i$, $recall(D,S_i)$ denotes the achieved recall by $S_i$, and $P(\cdot)$ denotes the probability of a required quality level being met.
\end{definition}

  Note that in Definition~\ref{optimization}, $P(\cdot)$, or the probability of satisfying a certain required quality level, is statistically equivalent to the confidence level $\theta$ defined in Definition.~\ref{problemsetting}. It can be observed that HUMO achieves a 100\% precision and recall levels in the extreme case when all the instance pairs are assigned to the human (i.e. $D_H$=$D$). In general, its achieved precision and recall levels tend to decrease as $D_H$ becomes smaller. However, the problem of searching for the minimum size $D_H$ is challenging due to the fact that the ground-truth match proportions of $D_-$ and $D_+$ are unknown. In the following sections, we propose three search approaches: a conservative baseline one purely based on the monotonicity assumption of precision (Section~\ref{sec:conservative}), a more aggressive sampling-based one (Section~\ref{sec:aggressive}), and a hybrid one that benefits from the strengths of both previous approaches (Section~\ref{sec:hybrid}). They estimate the match proportions of $D_-$ and $D_+$ based on different assumptions.

%% file: 3-baseline.tex
\section{Baseline Approach}\label{sec:conservative}

  The baseline approach assumes that the instance pairs in the workload of $D$ statistically satisfy monotonicity of precision. It begins with an initial medium similarity value (e.g. the boundary value of a classifier or simply a median value), and then incrementally identifies the upper and lower bounds of the similarity interval of $D_H$, $v^-$ and $v^+$.

\begin{figure}[!htb]
\setlength{\abovecaptionskip}{\figcaptionspace}
\centering
\subfigure[Incrementally moving the upper bound of $D_H$ right.]
{\includegraphics[width=\linewidth]{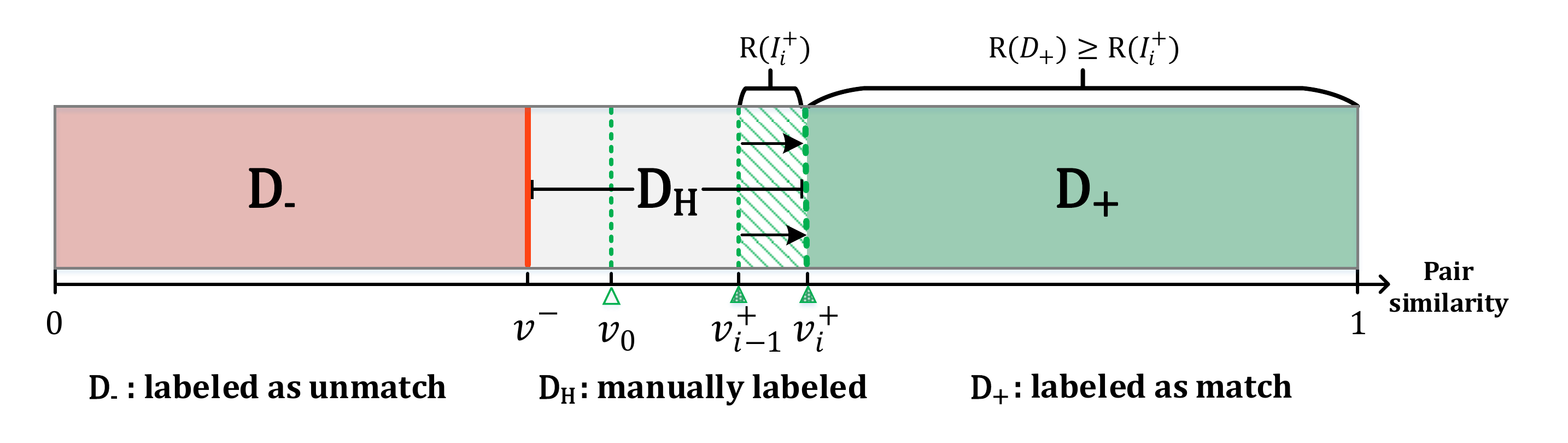}
\label{basic_idea_precision}}
\subfigure[Incrementally moving the lower bound of $D_H$ left.]
{\includegraphics[width=\linewidth]{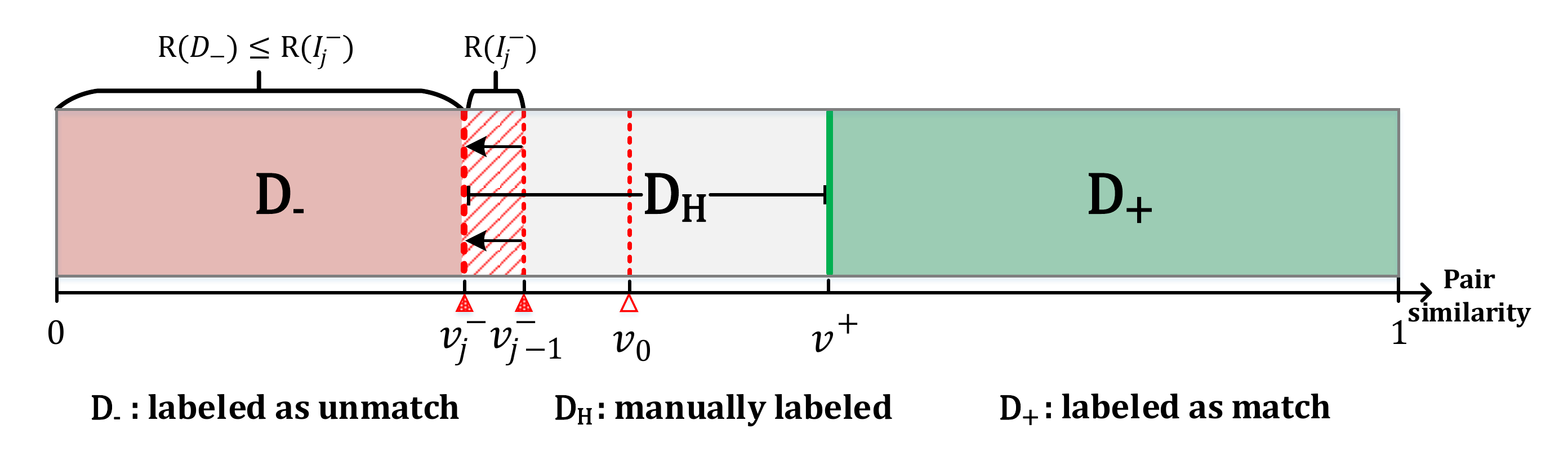}
\label{basic_idea_recall}}
\caption{The demonstration of the baseline solution.}
\end{figure}

   Initially, it sets $v^-$ and $v^+$ to a common value $v_0$, $v^-=v^+=v_0$. Then, it iteratively enlarges the similarity interval of $D_H$ until the desired precision and recall requirements are satisfied. Since both lower and upper bounds affect the precision and recall estimates, the search process alternately moves $v^-$ left and $v^+$ right.

  Suppose that $v^+$ is moved from $v_{i-1}^+$ to a higher value $v_i^+$, as shown in Figure \ref{basic_idea_precision}. It is clear that as the mark of $v^+$ is moved right, the number of true positives would remain constant while the number of false positives would decrease. As a result, the achieved precision would in turn increase. We denote the interval $(v_{i-1}^+,v_i^+]$ by $I_i^+$. According to the monotonicity assumption of precision, the match proportion of the pairs in the interval $(v_i^+,1]$ is no less than $\mathsf{R}(I_i^+)$, in which $\mathsf{R}(I_i^+)$ denotes the observed match proportion of the pairs in $I_i^+$. Therefore, with $v^-$ and $v^+=v_i^+$, the lower bound of the achieved precision level can be represented by
\begin{equation}
  \frac{|D_H|\cdot \mathsf{R}(D_H)+|D_+|\cdot \mathsf{R}(I_i^+)}{|D_H|\cdot \mathsf{R}(D_H)+|D_+|},
\end{equation}
in which $|D_H|$ and $|D_+|$ denote the total numbers of pairs in $D_H$ and $D_+$ respectively. Accordingly, given the precision requirement $\alpha$, the match proportion of the interval $I_i^+$ should satisfy
\begin{equation}
\mathsf{R}(I_i^+)\geq\frac{\alpha\cdot |D_+|-(1-\alpha)\cdot \mathsf{R}(D_H)\cdot |D_H|}{|D_+|}.
\label{eq:baseline-precision-condition}
\end{equation}
In other words, the precision requirement $\alpha$ would be satisfied once the observed match proportion of the interval $I_i^+$ reaches the threshold presented in Eq.~\ref{eq:baseline-precision-condition}.

  Similarly, suppose that the lower bound $v^-$ is moved from $v_{j-1}^-$ to a lower value $v_j^-$, as shown in Figure~\ref{basic_idea_recall}. We denote the interval $[v_j^-,v_{j-1}^-)$ by $I_j^-$. According to the monotonicity assumption of precision, the match proportion of the pairs in the interval $[0,v_j^-)$ is no larger than $\mathsf{R}(I_j^-)$. Therefore, with $v^+=v_i^+$ and $v^-=v_j^-$, the lower bound of the achieved recall level can be represented by
\begin{equation} \frac{|D_H|\cdot\mathsf{R}(D_H)+|D_+|\cdot\mathsf{R}(I_i^+)}{|D_-|\cdot\mathsf{R}(I_j^-)+|D_H|\cdot\mathsf{R}(D_H)+|D_+|\cdot\mathsf{R}(I_i^+)}.
\end{equation}
Accordingly, given the recall requirement $\beta$, the match proportion of the interval $I_j^-$ should satisfy
\begin{equation}
  \mathsf{R}(I_j^-)\leq\frac{(1-\beta)(|D_H|\cdot\mathsf{R}(D_H)+|D_+|\cdot\mathsf{R}(I_i^+))}{\beta\cdot |D_-|}.
\label{eq:baseline-recall-condition}
\end{equation}
In other words, the recall requirement $\beta$ would be satisfied once the observed match proportion of $I_j^-$ is below or equal to the threshold presented in Eq.~\ref{eq:baseline-recall-condition}.

	The search process alternately moves $v^+$ right and $v^-$ left to enforce precision and recall requirements. Once $\mathsf{R}(I_i^+)$ reaches the threshold specified in Eq.~\ref{eq:baseline-precision-condition}, the upper bound of $D_H$ would be finally fixed at $v_i^+$. It can be observed that with the upper bound fixed at $v_i^+$, moving $v^-$ to a lower value would only increase the estimated precision level. Similarly, once $\mathsf{R}(I_j^-)$ falls below the threshold specified in Eq.~\ref{eq:baseline-recall-condition}, the lower bound of $D_H$ would be finally fixed at $v_j^-$. Due to the monotonicity assumption, with the lower bound fixed at $v_j^-$, moving $v^+$ to a higher value would only increase the estimated recall level. In practical implementation, we can set the unit movement of $v^-$ and $v^+$ by the number of instance pairs: the intervals $(v_{i-1}^+,v_i^+]$ and $[v_j^-,v_{j-1}^-)$ always contain the same number of instance pairs. Further details on the search process are omitted here due to space limits, and a more thorough explanation can be referred to in our technical report \cite{chen2017humoreport}.

   By following the above reasoning, the baseline search process can return a solution satisfying the user-specified precision and recall levels with a 100\% confidence, provided that the monotonicity assumption holds. Its computational complexity is only linear with the number of instance pairs in $D$ in the worst case. Finally, we conclude this section with Theorem~\ref{theorem-baseline}, whose proof follows naturally from our above analysis.

\begin{theorem} \label{theorem-baseline}
  Given an ER workload of $D$, the baseline search process returns a HUMO solution that can ensure the precision and recall levels of $\alpha$ and $\beta$ respectively with the confidence of 100\% provided that the monotonicity assumption holds on $D$.
\end{theorem}

%% file: 3-sampling.tex
\section{Sampling-based Approach} \label{sec:aggressive}

  The baseline approach estimates the match proportions of $D_-$ and $D_+$ by the observed match proportions of the intervals in $D_H$. However, it can be noticeable that the match proportion of $D_-$ is usually significantly smaller than that of $D_H$, while the match proportion of $D_+$ is usually considerably larger than that of $D_H$. Strictly speaking, the baseline approach may overestimate the match proportion of $D_-$, and may also underestimate the match proportion of $D_+$. As a result, it would require considerably more than necessary human cost to enforce quality guarantees. To alleviate this limitation, we propose a more aggressive sampling-based approach in this section. Compared with the baseline approach, it is more aggressive in that it estimates the match proportions of $D_-$ and $D_+$ by directly sampling them.

  The sampling-based approach divides $D$ into multiple disjoint unit subsets and estimates their match proportions by sampling. We first present an all-sampling solution that samples all the subsets. To reduce human cost, we also present an improved partial-sampling solution that only requires sampling a portion of the subsets.

\begin{figure}[!htb]
\setlength{\abovecaptionskip}{\figcaptionspace}
\centering
\includegraphics[width=\linewidth]{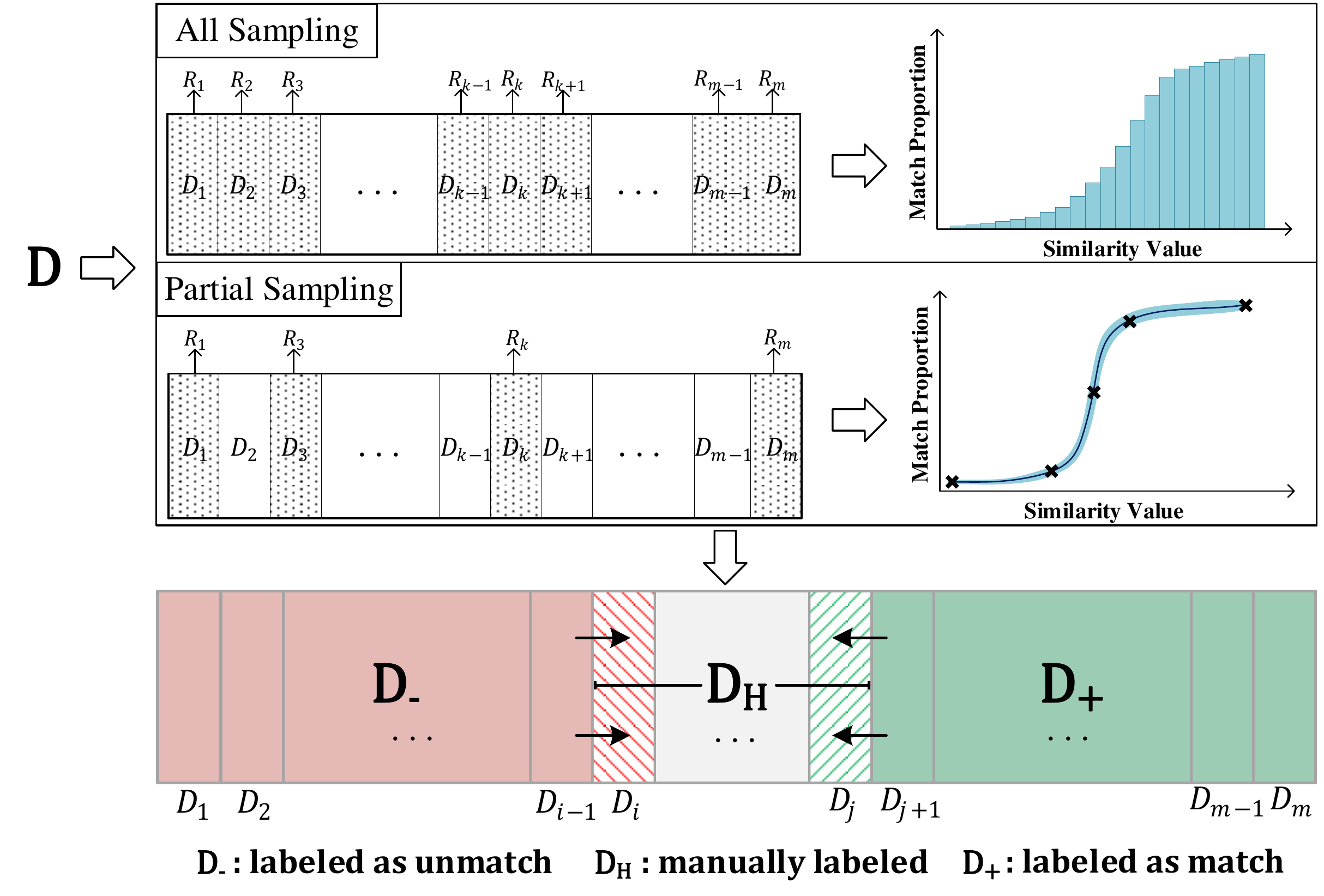}
\caption{The demonstration of sampling-based solution.}
\label{sampling_based_demonstration}
\end{figure}

\subsection{All-Sampling Solution} \label{sec:all-sampling}

   Suppose that $D$ is divided into $m$ disjoint subsets, $D=D_1\cup\cdots\cup D_m$, and the subsets are ordered by their similarity values. If $i<j$, then $\forall d\in D_i$ and $\forall d'\in D_j$, we have $sim(d)\leq sim(d')$, in which $sim(d)$ denotes the similarity value of $d$. With the notation of $D_i$, we can represent $D_H$ by a union of subsets, $D_H=D_i\cup D_{i+1}\cdots\cup D_j$, in which $D_i$ is the lower bound subset of $D_H$ while $D_j$ is its upper bound subset. We also denote the sampled match proportion of $D_i$ by $\mathsf{R}_i$. We first consider the hypothetical case that the estimate of $\mathsf{R}_i$ is accurate, and then integrate sampling errors into bound computation.

  When the estimate of $\mathsf{R}_i$ is hypothetically accurate, the achieved recall level of a HUMO solution solely depends on $D_H$'s lower bound. Therefore, the all-sampling solution first identifies $D_H$'s lower bound subset to meet the constraint on recall, then identifies its upper bound subset to meet the precision constraint. With the lower bound of $D_H$ set at $D_i$, the achieved recall level can be estimated by
\begin{equation}
\label{eq:all-recall}
  recall(D,S)=\frac{\sum_{i\leq k\leq m}{|D_k|\cdot \mathsf{R}_k}}{\sum_{1\leq k\leq m}{|D_k|\cdot \mathsf{R}_k}}.
\end{equation}
Therefore, to minimize the size of $D_H$ while ensuring the recall level $\beta$, the search process initially sets the lower bound to $D_1$, and then iteratively moves it right from $D_k$ to $D_{k+1}$ until the estimated recall level specified in Eq.~\ref{eq:all-recall} falls below $\beta$.

  The search process then deals with the precision constraint in a similar way by incrementally identifying the upper bound of $D_H$. Suppose that the lower bound of $D_H$ has been identified to be $D_i$. With its upper bound set at $D_j$, the achieved precision level can be estimated by
\begin{equation}
\label{eq:all-precision}
  precision(D,S)=\frac{\sum_{i\leq k\leq m}{|D_k|\cdot \mathsf{R}_k}}{\sum_{i\leq k\leq j}{|D_k|\cdot \mathsf{R}_k}+\sum_{j+1\leq k\leq m}{|D_k|}}
\end{equation}
Therefore, to minimize the size of $D_H$ while ensuring the precision level $\alpha$, the search process initially sets the upper bound to $D_m$, and then iteratively moves it left from $D_{k}$ to $D_{k-1}$ until the estimated precision level specified in Eq.~\ref{eq:all-precision} falls below $\alpha$.

  Now we describe how to integrate sampling errors into bound computation. For fulfilling the confidence level, we resort to the theory of stratified random sampling \cite{cochran1977sampling} to estimate sampling error margins. We denote the total number of pairs in $D$ by $n$ and the number of pairs in the subset $D_i$ by $n_i$. Based on the sampled match proportion estimate of $D_i$, we can compute the mean of the match proportion of $D$ and its estimated standard deviation, which are denoted by ${\bar{\mathsf{R}}_D}$ and $\sigma_D$ respectively. The details on how to compute ${\bar{\mathsf{R}}_D}$ and $\sigma_D$ can be found in~\cite{chen2017humoreport}. Given the confidence level $\theta$, the total number of matching pairs in $D$ falls reasonably within the interval
\begin{equation}
  [n\cdot(\bar{\mathsf{R}}_D-t_{(1-\theta, d.f.)}\cdot\sigma_{D}), n\cdot(\bar{\mathsf{R}}_D+t_{(1-\theta, d.f.)}\cdot\sigma_{D})],
\label{eq:confidenceintervals}
\end{equation}
in which $t_{(1-\theta, d.f.)}$ is {\em Student's t value} for {\em d.f.} degrees of freedom and the confidence level $\theta$ for two-sided critical regions. In Eq.~\ref{eq:confidenceintervals}, as typical in stratified sampling \cite{cochran1977sampling}, we use Student's t value to take account of the sampling error due to limited sample size. Suppose that a random variable $T$ has a Student's t-distribution, a Student's t value for confidence level $\theta$ for two-sided critical regions is the value, let's say $\tilde{t}$, that satisfies $P(-\tilde{t} < T < \tilde{t})=\theta$, where $P(\cdot)$ represents the probability.

  Next, we apply the analysis results of confidence error margins in the recall and precision estimates as presented in Eq.~\ref{eq:all-recall} and ~\ref{eq:all-precision}. According to Eq.~\ref{eq:all-recall}, the lower bound of the recall estimate can be guaranteed by setting a lower bound on $n_{[i,m]}^{+}$ and an upper bound on $n_{[1,i-1]}^{+}$, in which $n_{[i,j]}^{+}$ denotes the total number of matching pairs in the subset union, $D_i\cup D_{i+1}\cdots\cup D_j$. Suppose that the lower bound of $D_H$ is set at $D_i$. Given the confidence level $\theta$ and the recall level $\beta$, the HUMO solution meets the recall requirement if
\begin{equation}
\label{eq:recall-condition}
  \beta\leq \frac{lb(n_{[i,m]}^{+}, \sqrt{\theta})}{ub(n_{[1,i-1]}^{+}, \sqrt{\theta}) + lb(n_{[i,m]}^{+}, \sqrt{\theta})},
\end{equation}
in which $lb(n_{[i,m]}^{+}, \sqrt{\theta})$ denotes the lower bound of $n_{[i,m]}^{+}$ with the confidence $\sqrt{\theta}$, and $ub(n_{[1,i-1]}^{+}, \sqrt{\theta})$ denotes the upper bound of $n_{[1,i-1]}^{+}$ with the confidence $\sqrt{\theta}$. Since the bound estimations on $n_{[i,m]}^{+}$ and $n_{[1,i-1]}^{+}$ are independent, the lower bound of the recall level specified in Eq.~\ref{eq:recall-condition} has the desired confidence $\theta$.

  Similarly, suppose that the lower and upper bounds of $D_H$ are set at $D_i$ and $D_j$ respectively. Given the confidence level $\theta$ and the precision level $\alpha$, the HUMO solution meets the precision requirement if
\begin{equation}
\label{eq:precision-condition}
  \alpha\leq \frac{lb(n_{[i,j]}^{+}, \sqrt{\theta}) + lb(n_{[j+1,m]}^{+}, \sqrt{\theta})}{lb(n_{[i,j]}^{+}, \sqrt{\theta}) + n_{[j+1,m]}}.
\end{equation}
Since the bound estimations on $n_{[i,j]}^{+}$ and $n_{[j+1,m]}^{+}$ are independent, the lower bound of the precision level specified in Eq.~\ref{eq:precision-condition} has the desired confidence $\theta$.

  The all-sampling search process iteratively searches for the lower and upper bounds of $D_H$. It first identifies the maximal value of $i$ such that the condition specified in Eq.~\ref{eq:recall-condition} is satisfied. It begins with $i=1$ and then iteratively moves the lower bound right from $D_i$ to $D_{i+1}$. Similarly, with the lower bound of $D_H$ set at $D_i$, it identifies the minimal value of $j$ such that the condition specified in Eq.~\ref{eq:precision-condition} is satisfied. It begins with $j=m$ and then iteratively moves the upper bound left from $D_j$ to $D_{j-1}$. More details on the search process are however omitted here due to space constraints. They can be found in our technical report \cite{chen2017humoreport}.

  The worst-case computational complexity of the all-sampling search process can be represented by ${\bf O}(n+m^2)$, in which $n$ denotes the total number of pairs in $D$ and $m$ denotes the total number of subsets. Finally, we conclude this subsection with the following theorem, whose proof follows naturally from our above analysis:
\begin{theorem}
  Given an ER workload of $D$, a confidence level $\theta$, a precision level $\alpha$ and a recall level $\beta$, the all-sampling search process returns a HUMO solution that can ensure the precision and recall levels of $\alpha$ and $\beta$ respectively with the confidence $\theta$.
\end{theorem}

\subsection{Partial-Sampling Solution} \label{sec:partial-sampling}

 Note that samples should be labeled by the human and the optimization objective of HUMO is to minimize human cost. The all-sampling solution has to sample every subset; therefore its human cost consumed on labeling samples is usually prohibitive. In this subsection, we propose an improved solution that only needs to sample a portion of the subsets. It achieves the purpose by approximating the match proportions of unsampled subsets based on those observed on sampled ones. We use the Gaussian process (GP) \cite{rasmussen2006gaussian}, which is a classical technique for non-parametric regression. GP assumes that the match proportions of subsets have a joint Gaussian distribution. It can smoothly integrate sampling error margins into the approximation process.

   Given $k$ sampled subsets, we denote their observed match proportions by $\mathsf{R} = [\mathsf{R}_1, \mathsf{R}_2, \ldots, \mathsf{R}_k]^T$, and their corresponding average similarity values by $V = [v_1, v_2, \ldots, v_k]^T$. The Gaussian process estimates the match proportion, $\mathsf{R}_*$, of a new similarity value, $v_*$, based on $\mathsf{R}$, the observed match proportions of $V$. According to the assumption of GP, the random variables $[V^T, v_*]^T$ satisfy a joint Gaussian distribution, which can be represented by
  \begin{equation}
  \begin{bmatrix} V \\ v_* \end{bmatrix} \sim
  \mathcal{N}\left(0, \begin{bmatrix} \mathbf{K}(V, V) & \mathbf{K}(V, v_*) \\ \mathbf{K}(v_*, V) & \mathbf{K}(v_*, v_*) \end{bmatrix}\right),
  \label{eq:jointdistribution}
  \end{equation}
in which $\mathbf{K}(\cdot , \cdot)$ represents the covariance matrix. The details of how to compute the covariance matrix $\mathbf{K}(\cdot , \cdot)$ can be found in~\cite{chen2017humoreport}. Based on Eq.~\ref{eq:jointdistribution}, the mean of the match proportion of $v_*$, $\mathsf{R}_*$, can be represented by
  \begin{equation}
  \bar{\mathsf{R}}_* = \mathbf{K}(v_*, V) \cdot \mathbf{K}^{-1}(V, V) \cdot \mathsf{R}.
  \label{eq:gpr:mean}
  \end{equation}
The variance of $\mathsf{R}_*$ can be also represented by
\begin{equation}
  \sigma_{\mathsf{R}_*}^2 = \mathbf{K}(v_*, v_*) - \mathbf{K}(v_*, V) \cdot \mathbf{K}^{-1}(V, V) \cdot \mathbf{K}(V, v_*).
\label{eq:gpr:variance}
\end{equation}
Accordingly, the distribution of $\mathsf{R}_*$, the match proportion of $v_*$, can be represented by the following Gaussian function
\begin{equation}
  \mathsf{R}_* \sim \mathcal{N}\left(\bar{\mathsf{R}}_*, \sigma_{\mathsf{R}_*}^2\right).
\end{equation}

  Now we are ready to describe how to aggregate the estimations of multiple subsets. Note that the distribution of each subset's match proportion satisfies a Gaussian function. Given the $t$ subsets of $D_*$, $D_*$ = $\{D_*^1,D_*^2,\ldots,D_*^t\}$, we denote their corresponding numbers of pairs by $\{n_*^1, n_*^2,\ldots, n_*^t\}$, and their similarity values by $V_*=[v_*^1, v_*^2, \ldots, v_*^t]^T$. Then, the total number of match pairs in $D_*$, denoted by $n_*$, satisfies a Gaussian distribution. Its mean can be represented by
\begin{equation}
  \bar{n}_* = \sum_{i=1}^{t}n_*^i\cdot\bar{\mathsf{R}}_*^i,
\end{equation}
in which $\bar{\mathsf{R}}_*^i$ represents the mean of the match proportion of $D_*^i$. Its standard deviation can also be represented by
\begin{equation}
  \sigma_{D_*} = \sqrt{\sum_{1\leq i\leq t,1\leq j\leq t}n_*^i\cdot n_*^j\cdot cov(v_*^i, v_*^j)},
\end{equation}
in which $cov(v_*^i, v_*^j)$ is the covariance between two estimates and its value is the {\em (i,j)-th} element in the covariance matrix $\mathbf{K}(V_*,V_*)-\mathbf{K}(V_*,V)\cdot\mathbf{K^{-1}}(V,V)\cdot\mathbf{K}(V,V_*)$. Therefore, given the confidence level $\theta$, the corresponding confidence interval of the number of match pairs in $D_*$ can be represented by
\begin{equation}
  [\bar{n}_* - \mathcal{Z}_{(1-\theta)} \cdot \sigma_{D_*}, \bar{n}_* + \mathcal{Z}_{(1-\theta)} \cdot \sigma_{D_*}],
\label{eq:gpr:confidenceintervals}
\end{equation}
in which $\mathcal{Z}_{(1-\theta)}$ is the $(1-\frac{1-\theta}{2})$ point of {\em standard normal distribution}.

\begin{algorithm}
\setlength{\textfloatsep}{0pt}
\caption{Gaussian Regression of Match Proportion Function}
\label{alg:fit-gp}
\KwIn{Sorted disjoint subsets $\{D_1, D_2, ..., D_m\}$; Sampling cost range $[p^l, p^u]$; Error threshold $\varepsilon$.}
\KwOut{The function of match proportion, $F_k$.}
$j \gets m\cdot p^l$\;
$TrainSet \gets$ select j equidistance subsets $\{D_{i_1},$ $D_{i_2},$ $...,$ $D_{i_j}\}$\;
$\mathsf{V}, \mathsf{R} \gets$ sample every subset in $TrainSet$ to get their match proportion estimates\;
$F_k \gets$ use $\mathsf{V}, \mathsf{R}$ to train Gaussian process model\;
$IndexQueue \gets [(i_1, i_2), ..., (i_k, i_{k+1}), ..., (i_{j-1}, i_j)]$\;
\While{$IndexQueue$ is not empty \\ \qquad and $|TrainSet| < m\cdot p^u$}
{
    $(i_k, i_{k+1}) \gets IndexQueue.pop()$\;
    $D_x \gets$ the middle subset between $D_{i_k}$ and $D_{i_{k+1}}$\;
    $\mathsf{R}_x \gets$ match proportion of $D_x$ estimated by sampling\;
    \If{$|F_k(v_x) - \mathsf{R}_x| \geq \varepsilon$}
    {
        $IndexQueue.append([(i_k, x),(x, i_{k+1})])$\;
    }
    Add $D_x, v_x, \mathsf{R}_x$ to $TrainSet, \mathsf{V}, \mathsf{R}$ respectively\;
    $F_k \gets$ use $\mathsf{V}, \mathsf{R}$ to train Gaussian process model\;
}
return $F_k$.
\end{algorithm}

  The partial-sampling search process consists of two phases. It trains the function of match proportion by Gaussian regression in the first phase, it then searches for the lower and upper bounds of $D_H$ based on the trained function in the second phase. The function training's procedure is sketched in Algorithm\ref{alg:fit-gp}. Note that $D$ is divided into $m$ disjoint subsets \{$D_1$, $D_2$, $\ldots$, $D_m$\}. To balance approximation accuracy and sampling cost, it presets a range, $[p^l, p^u]$ (e.g. $[1\%, 5\%]$), for the proportion of sampled subsets among all subsets. Initially, the training set consists of $j$ sampled subsets, \{$D_{i_1}$, $D_{i_2}$, $\ldots$, $D_{i_j}$\}, in which $j=m\times p^l$ and $\forall 1\leq k\leq j-2$, $i_{k+1}-i_k=i_{k+2}-i_{k+1}$. In each iteration, the algorithm first trains an approximation function, denoted by $F_k$, by Gaussian regression based on the sampled subsets. It then uses $F_k$ to estimate the match proportion of a subset that is located in the middle point between two neighbouring sampled subsets. Suppose that $D_x$ denotes the subset between the sampled subsets $D_{i_k}$ and $D_{i_{k+1}}$. If the difference between the estimated value based on $F_k$ and the observed match proportion based on sampling exceeds a small threshold $\epsilon$, the algorithm would add $D_x$ into the training set; otherwise, it would not sample any other subset between $D_{i_k}$ and $D_{i_{k+1}}$ (except $D_x$) in the following iterations. Finally, the algorithm trains the function with the updated training set. This cycle of sampling and training is iteratively invoked until the trained function achieves a good approximation or the sampling cost reaches the upper bound of the pre-specified range (i.e. $p^u$).

  Similar to the procedure for all-sampling solution, the partial-sampling search process first identifies the maximal lower bound of $D_H$ to meet the recall requirement, and then identifies the minimal upper bound of $D_H$ to meet the precision requirement. The only difference is that the lower bounds of the achieved recall and precision levels of a HUMO solution should be estimated by the confidence intervals specified in Eq.~\ref{eq:gpr:confidenceintervals}.

  The worst-case computational complexity of Alg.~\ref{alg:fit-gp} is in the order of ${\bf O}(k^4)$, in which $k$ denotes the number of sampled unit subsets. The worst-case computational complexity of the search process can be represented by ${\bf O}(m\cdot k^2+m^3)$. Therefore, the worst-case computational complexity of the partial-sampling solution can be represented by ${\bf O}(n+m^3+m\cdot k^2+k^4)$. It can be observed that the effectiveness of the partial-sampling solution in ensuring quality guarantees depends on the accuracy of the Gaussian approximation. As shown by our empirical evaluation in Section~\ref{sec:experiment}, the partial-sampling solution is highly effective due to the powerfulness and robustness of the Gaussian process.

%% file: 3-hybrid.tex
\section{Hybrid Approach}\label{sec:hybrid}

  The baseline approach usually overestimates the match proportion of $D_-$ while underestimating that of $D_+$. The sampling-based approach can alleviate both drawbacks to a large extent by directly sampling $D_-$ and $D_+$. However, it still has to consider confidence margins in the estimations of $D_-$ and $D_+$. Furthermore, it usually cannot afford to sample all the subsets in $D_-$ and $D_+$ due to prohibitive sampling cost. Generally, less samples would result in larger error margins. Therefore, there is no guarantee that a sampling-based estimation would always be better than the corresponding baseline one. As we show in Section~\ref{sec:experiment}, their relative performance actually depends on the characteristics of the given ER workload. This observation motivates us to propose a hybrid approach, which can take advantage of both estimations and use the better of both worlds in the process of bound computation.

  The hybrid approach begins with a HUMO solution of the partial-sampling approach. We denote the initial solution by $S_0$ and its lower and upper bounds of $D_H$ by $D_i$ and $D_j$ respectively. It searches for a better solution than $S_0$ by incrementally redefining $D_H$'s bounds using the better between the baseline and sampling-based estimates. Initially, it sets $D_H$ to be the single median subset of $D_i$ and $D_j$, $D_{\frac{i+j}{2}}$. Similar to the baseline approach, it alternately extends $D_H$'s upper and lower bounds until both precision and recall requirements are met. However, on reasoning about the match proportions of $D_-$ and $D_+$, instead of being purely based on the monotonicity of precision, it uses the better of both estimates. It alternately moves the upper bound from $D_u$ to $D_{u+1}$ and the lower bound from $D_l$ to $D_{l-1}$. After each movement of the upper bound, it checks whether the current solution satisfies the precision requirement. Similarly, after each movement of the lower bound, it checks whether the current solution satisfies the recall requirement. Note that the new range of $D_H$ can not exceed the range of $[D_i, D_j]$ in the initial solution $S_0$. Therefore, the resulting HUMO solution of the hybrid approach is at least as good as $S_0$. The details of the hybrid search process are omitted here due to space limits, but can be found in our technical report \cite{chen2017humoreport}.

  The worst-case computational complexity of the hybrid solution is the same as that of the partial-sampling solution, bounded by ${\bf O}(n+m^3+m\cdot k^2+k^4)$. Its effectiveness in ensuring quality guarantees depends on both the monotonicity assumption of precision and the accuracy of Gaussian approximation. As shown by our empirical evaluation in Section~\ref{sec:experiment}, the hybrid solution is highly effective in ensuring quality guarantees for HUMO.

%% file: 4-experiments.tex
\section{\vspace{-0.1cm}Experimental Evaluation} \label{sec:experiment}

  This section empirically evaluates HUMO's performance by performing a comparative study. We have implemented three proposed optimization approaches for HUMO:
\begin{itemize}
\item Baseline (denoted by BASE). It represents the optimization approach purely based on the monotonicity assumption of precision presented in Section~\ref{sec:conservative};
\item Sampling-based (denoted by SAMP). Since the all-sampling solution performs worse than the partial-sampling one, SAMP represents the partial-sampling solution presented in Section~\ref{sec:partial-sampling}. The comparative evaluation results between the all-sampling and partial-sampling solutions can be found in technical report~\cite{chen2017humoreport}.
\item Hybrid  (denoted by HYBR). It represents the hybrid approach presented in Section~\ref{sec:hybrid}.
\end{itemize}

   		In all HUMO's implementations, we divide an ER workload $D$ into disjoint subsets, each of which contains the same number of instance pairs. The number of instance pairs contained by each subset is set to be 200. Due to the distribution irregularity of matching pairs, BASE estimates the match proportion bounds of $D_-$ and $D_+$ by using the average match proportion of multiple consecutive subsets in $D_H$ instead of a single one. For practical implementation, we suggest that the number of consecutive subsets should be set between 3 and 10. Note that as its value increases, BASE becomes more conservative. To balance sampling cost and accuracy of match proportion approximation, SAMP sets both lower and upper limits on sampling cost, which is measured by the proportion of sampled subsets among all the subsets. In our implementation, the sampling proportion range is set to be between 1\% and 5\%.

 Note that most of the existing techniques for ER cannot enforce quality guarantees. Therefore, we implement a classical technique based on Support Vector Machine (SVM) \cite{kopcke2010evaluation} and present its results for quality reference. More recently proposed techniques (e.g. \cite{kouki2017collective} and \cite{lacoste2013sigma}) may have achieved better performance. However, it should be clear that as the SVM-based technique, they can not enforce quality guarantees. The performance difference between the existing classification techniques is beyond the scope of this paper. We instead compare HUMO with state-of-the-art active learning based approaches \cite{arasu2010active, bellare2012active} (denoted by ACTL), which can at least enforce precision. ACTL maximizes recall level while ensuring a user-specified precision level. It estimates the achieved precision level of a given labeling solution by sampling. It also requires manual verification. We compare HUMO and ACTL on achieved quality and required human cost.

\begin{figure}
\setlength{\abovecaptionskip}{\figcaptionspace}
\centering
\subfigure[DS Dataset.]
{\includegraphics[width=0.48\linewidth]{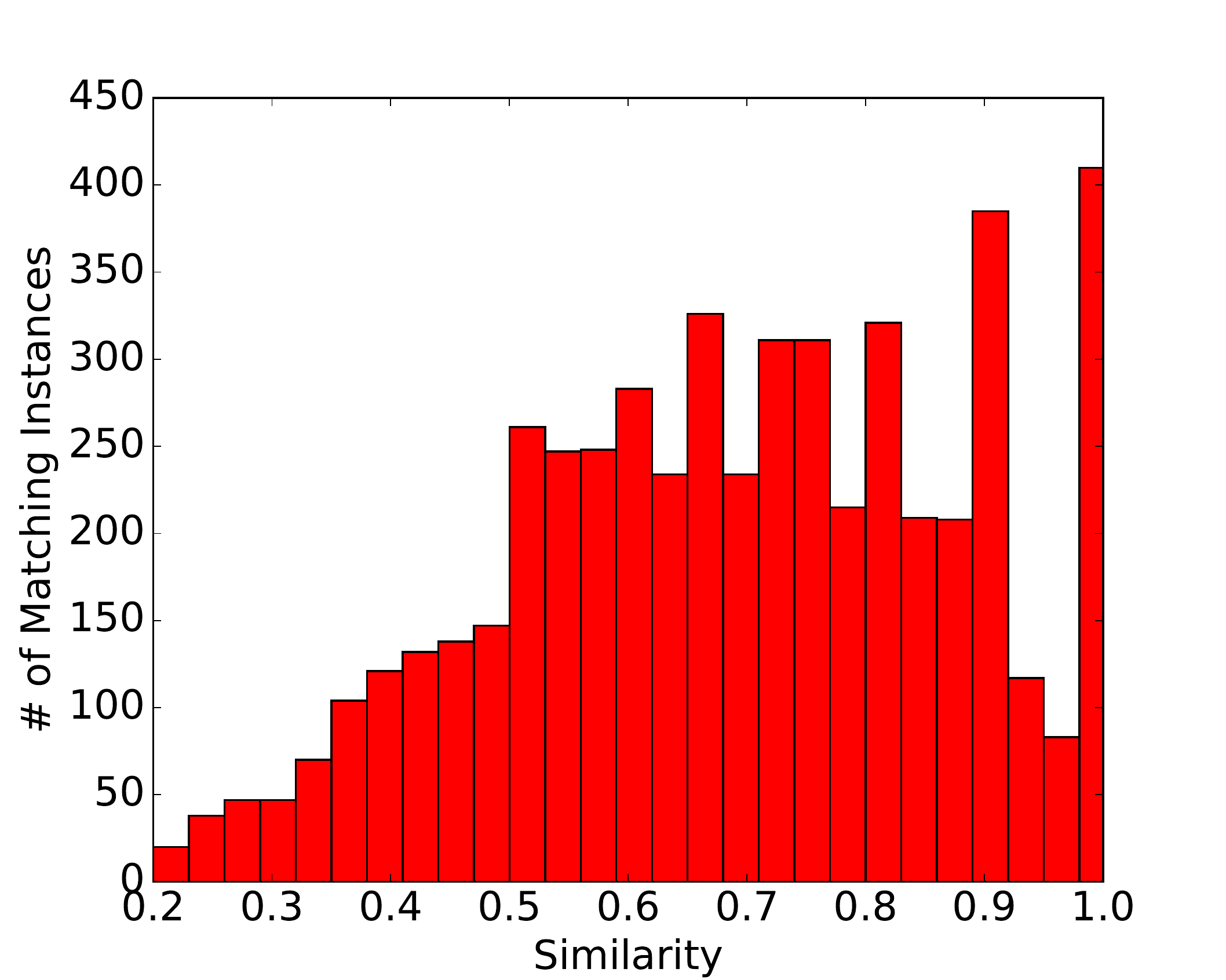}}
\subfigure[AB Dataset.]
{\includegraphics[width=0.48\linewidth]{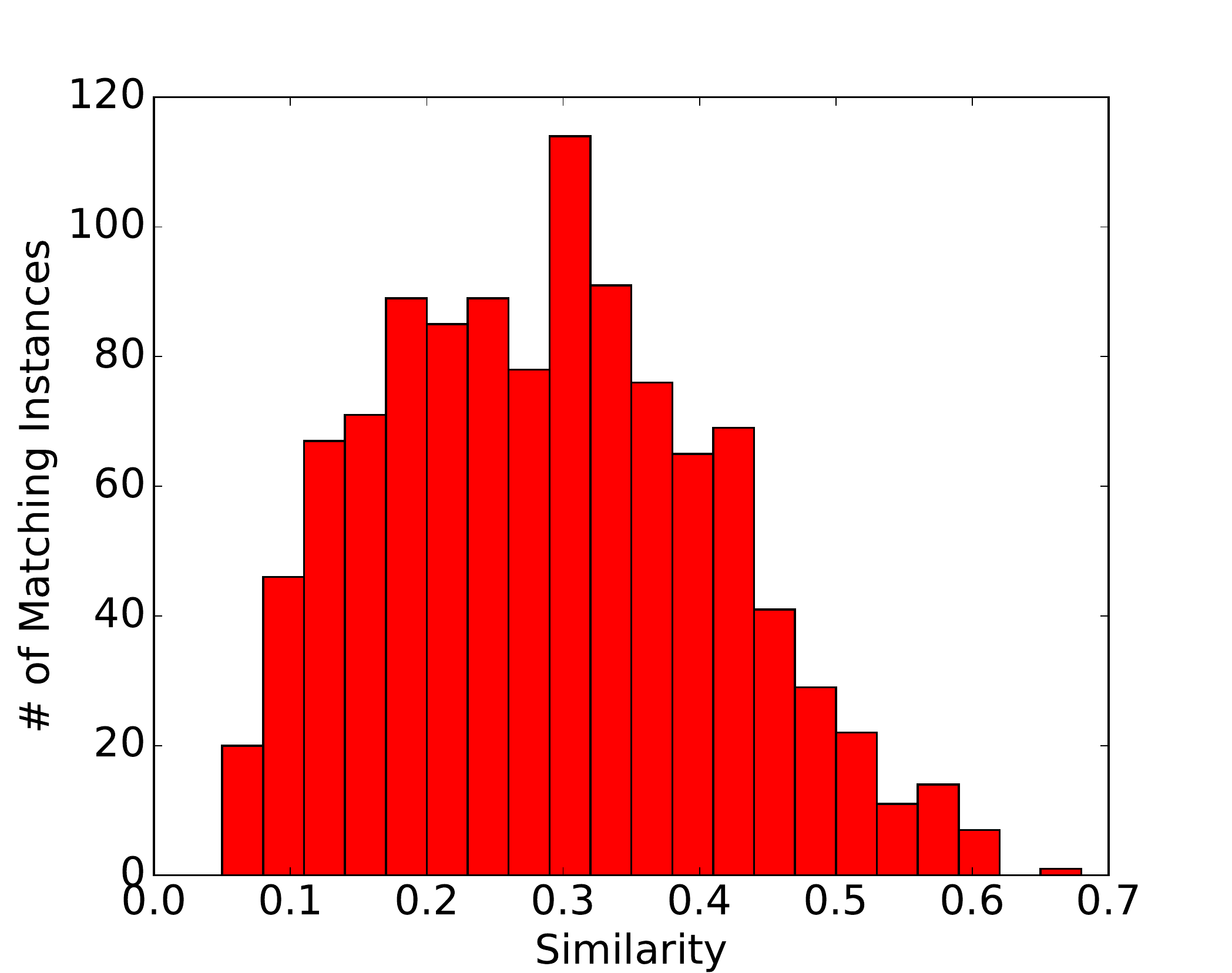}}
\caption{The distributions of matching pairs in two real datasets.}
\label{fig:matching-distribution}
\vspace{-0.8cm}
\end{figure}

  The rest of this section is organized as follows: Subsection~\ref{sec:datasets} describes our experimental setup. Subsection~\ref{sec:HUMO-experiment} evaluates the performance of different optimization approaches for HUMO. Subsection~\ref{sec:other-comparison} compares HUMO with ACTL. Finally, Subsection~\ref{sec:scalability} evaluates its efficiency and scalability.
\vspace{-0.25cm} 
\subsection{\vspace{-0.15cm}Experimental Setup}
\label{sec:datasets}

  We use two real datasets and one synthetic dataset in our evaluation. The experiments on real datasets can demonstrate the proposed solutions' performance in real application scenarios. The experiments on synthetic datasets can instead test their performance sensitivity to different data characteristics. The details of the two real datasets~\cite{kopcke2010evaluation} are described as follows:
\begin{itemize}
\item DBLP-Scholar\footnote{\url{https://dbs.uni-leipzig.de/file/DBLP-Scholar.zip}} (denoted by DS): The DBLP dataset contains 2616 publication entities from DBLP publications and the Scholar dataset contains 64263 publication entities from Google Scholar. The experiments match the DBLP entries with the Scholar entries.
\item Abt-Buy\footnote{\url{https://dbs.uni-leipzig.de/file/Abt-Buy.zip}} (denoted by AB): It contains 1081 product entities from Abt.com and 1092 product entities from Buy.com. The experiments match the Abt entries with the Buy entries.
\end{itemize}

  On both datasets, we compute pair similarity by aggregating attribute similarities with weights \cite{christen2012data}.  Specifically, on the DS dataset, Jaccard similarity of the attributes {\em title} and {\em authors}, and Jaro-Winkler distance of the attribute {\em venue} are used; on the AB dataset, Jaccard similarity of the attributes {\em product name} and {\em product description} are used. The weight of each attribute is determined by the number of its distinct attribute values. As in \cite{arasu2010active}, we use the blocking technique to filter the instance pairs unlikely to match. Specifically, the workload of DS contains the instance pairs whose aggregated similarity values are no less than 0.2. Similarly, the aggregated similarity value threshold for the AB workload is set to be 0.05. After blocking, the DS dataset has 100077 pairs and 5267 among them are matching pairs; the AB dataset has 313040 pairs and 1085 among them are matching pairs.

 The distributions of matching pairs in the two real datasets are presented in Figure~\ref{fig:matching-distribution}, in which the X-axis represents pair similarity value and the Y-axis represents the number of matching pairs. It can be observed that in DS, the majority of matching pairs has high similarity values; in AB, many matching pairs however have median and low similarity values. Therefore, in terms of classification accuracy, AB is a more challenging workload than DS.

   The performance of the SVM-based technique on the metrics of precision, recall and F1 are presented in Table~\ref{tab:svm-results}. Note that similar results have also been reported in \cite{kopcke2010evaluation}. However, the performance of the SVM-based solution is highly dependent on the selected features and training data. Here we only use them for quality reference. It can be observed that the classification quality on DS is better than that on AB. This observation is consistent with the two datasets' matching pairs distributions presented in Figure~\ref{fig:matching-distribution}.

    \begin{table}
    \caption{The SVM-based classification results on DS and AB.}
    \vspace{-0.1cm}
    \centering
    \label{tab:svm-results}
    \begin{tabular}{|c|c|c|c|}
    \hline
    Dataset & Precision & Recall & F1 Score \\
    \hline
    DS & 0.87 & 0.76 & 0.81 \\
    \hline
    AB & 0.47 & 0.35 & 0.40 \\
    \hline
    \end{tabular}
    \vspace{-0.1cm}
    \end{table}

\begin{figure}
\setlength{\abovecaptionskip}{\figcaptionspace}
\centering
{\includegraphics[width=0.55\linewidth]{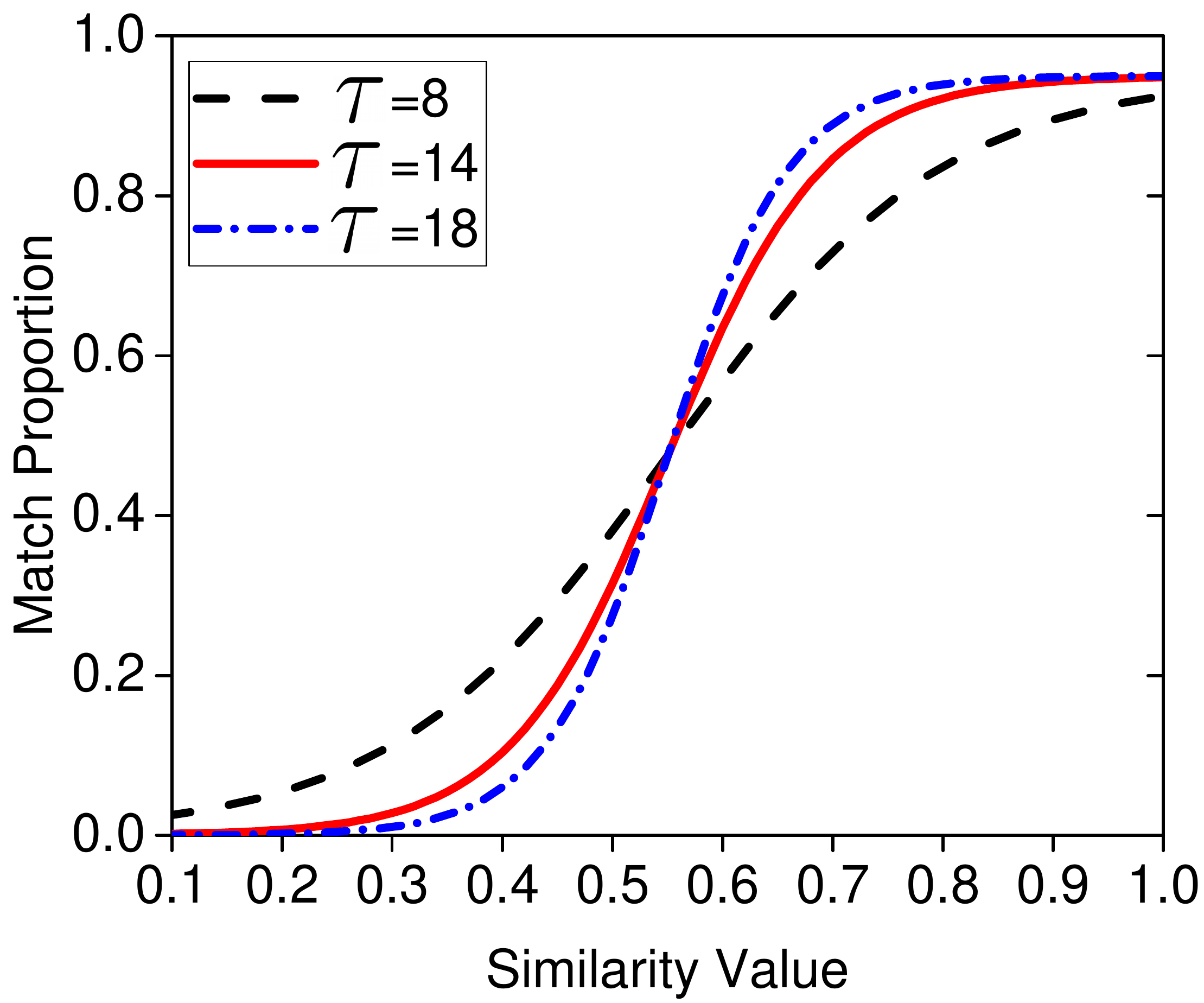}}
\caption{Logistic function.}
\label{fig:logistic-func}
\vspace{-0.65cm}
\end{figure}

  The generator for synthetic datasets uses the logistic function to simulate the function of match proportion with regard to pair similarity. The logistic function is represented by
\begin{equation}
  \frac{0.95}{1+e^{(-\tau(v-0.55))}}
\end{equation}
in which $v$ denotes pair similarity and the parameter $\tau$ specifies the steepness of the logistic curve. Some examples of the logistic function are also shown in Figure~\ref{fig:logistic-func}. As the value of $\tau$ decreases, the curve becomes less steep; the generated ER workload would be more challenging. The generator also has the parameter $\sigma$, which specifies the variances of the subsets' match proportions. A larger value of $\sigma$ would result in more distribution irregularity; the generated ER workload would be more challenging.

  Note that in our experiments, we have the ground-truth labels for all the test pairs. The ground-truth labels are originally hidden; whenever manual verification is called for, they are provided to the program. Existing crowdsourcing platforms can obviously be used to perform manual verification. Integrating HUMO with crowdsourcing platforms is an interesting future work. It is however beyond the scope of this paper.
	
  On efficiency evaluation, all experiments were run on a commercial machine running Windows 10, equipped with an Intel Core i5 2.30GHz and 16GB of RAM.
	
\subsection{Evaluation of HUMO Optimization} \label{sec:HUMO-experiment}

\subsubsection{On Real Datasets}

\begin{figure}
\setlength{\abovecaptionskip}{\figcaptionspace}
\centering
\subfigure[DS dataset.]
{\includegraphics[width=0.48\linewidth]{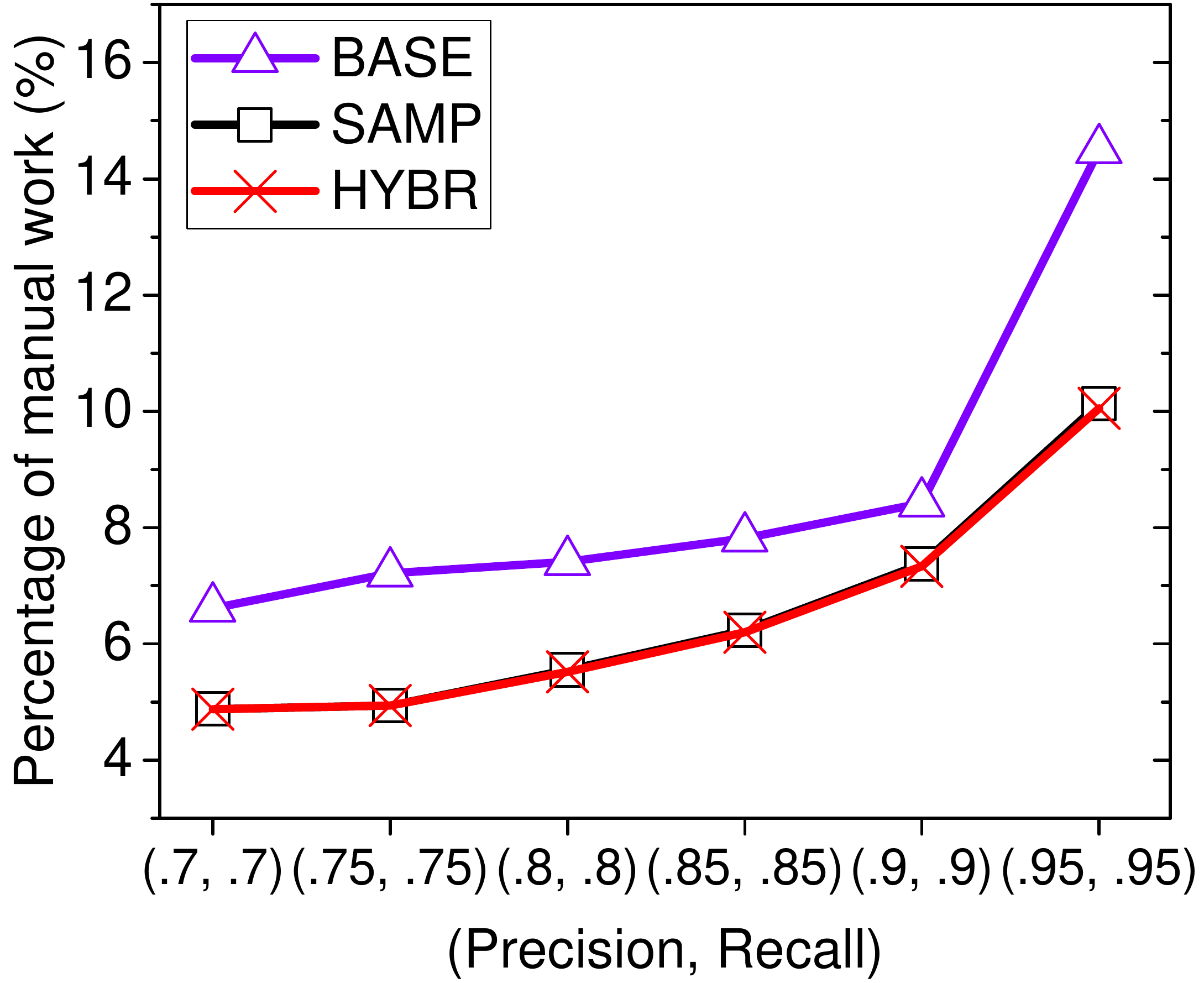}}
\subfigure[AB dataset.]
{\includegraphics[width=0.48\linewidth]{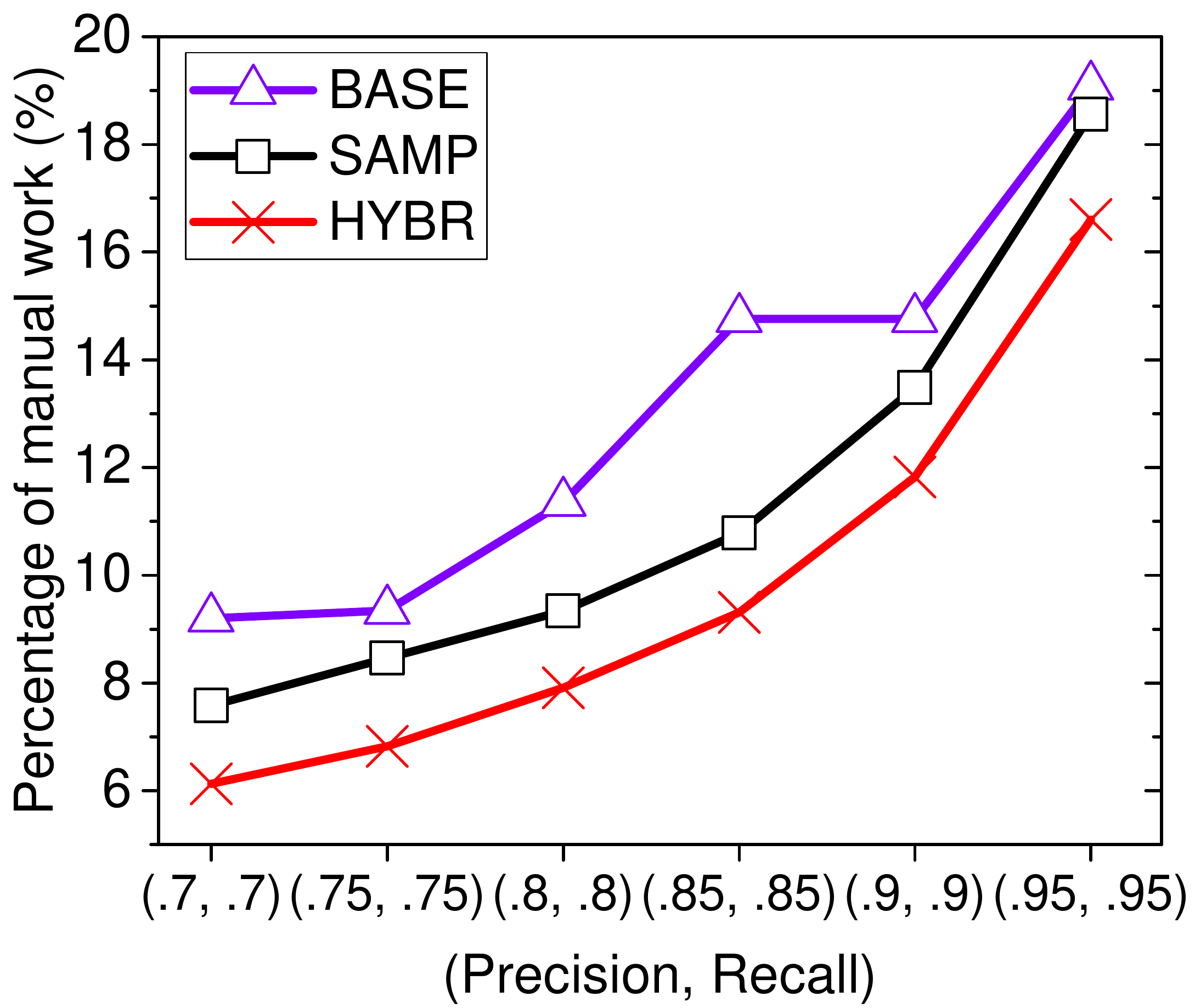}}
\caption{Comparison of human cost on two real datasets (with $\theta =0.9$).}
\label{fig:realdata-evaluation}
\end{figure}

\begin{table}
\vspace{-0.2cm}
\caption{The quality levels achieved by BASE on DS and AB.}
\vspace{-0.1cm}
\centering
\label{tab:con-quality}
\begin{tabular}{|c|c|c|}
\hline
Quality & \multicolumn{2}{c|}{Quality Levels of Results} \\
\cline{2-3}
Requirement & DS & AB \\
\hline
$\alpha=0.70$ & $\bar{\alpha}=0.9679$ & $\bar{\alpha}=0.9843$ \\
$\beta=0.70$ & $\bar{\beta}=0.9725$ & $\bar{\beta}=0.9244$ \\
\hline
$\alpha=0.75$ & $\bar{\alpha}=0.9732$ & $\bar{\alpha}=0.9843$ \\
$\beta=0.75$ & $\bar{\beta}=0.9738$ & $\bar{\beta}=0.9244$ \\
\hline
$\alpha=0.80$ & $\bar{\alpha}=0.9786$ & $\bar{\alpha}=0.9845$ \\
$\beta=0.80$ & $\bar{\beta}=0.9738$ & $\bar{\beta}=0.9382$ \\
\hline
$\alpha=0.85$ & $\bar{\alpha}=0.9786$ & $\bar{\alpha}=1.0\ \ \ \ $ \\
$\beta=0.85$ & $\bar{\beta}=0.9744$ & $\bar{\beta}=0.9521$ \\
\hline
$\alpha=0.90$ & $\bar{\alpha}=0.9883$ & $\bar{\alpha}=1.0\ \ \ \ $ \\
$\beta=0.90$ & $\bar{\beta}=0.9744$ & $\bar{\beta}=0.9521$ \\
\hline
$\alpha=0.95$ & $\bar{\alpha}=0.9946$ & $\bar{\alpha}=1.0\ \ \ \ $ \\
$\beta=0.95$ & $\bar{\beta}=0.9852$ & $\bar{\beta}=0.9659$ \\
\hline
\end{tabular}
\vspace{-0.5cm}
\end{table}

\begin{table}
\vspace{-0.4cm}
\caption{The quality levels achieved by SAMP on DS and AB.}
\vspace{-0.1cm}
\centering
\label{tab:agg-quality}
\begin{tabular}{|c|c|c|c|c|}
\hline
Quality & \multicolumn{2}{c|}{Quality Levels of Results} & \multicolumn{2}{c|}{Success rate}\\
\cline{2-5}
Requirement & DS & AB  & \ DS\ \  & AB  \\
\hline
$\alpha=0.70$ & $\bar{\alpha}=0.8649$ & $\bar{\alpha}=0.9282$ & \multirow{2}{*}{100} & \multirow{2}{*}{100} \\
$\beta=0.70$ & $\bar{\beta}=0.8365$ & $\bar{\beta}=0.8849$ & & \\
\hline
$\alpha=0.75$ & $\bar{\alpha}=0.8347$ & $\bar{\alpha}=0.9597$ & \multirow{2}{*}{100} & \multirow{2}{*}{100}\\
$\beta=0.75$ & $\bar{\beta}=0.8574$ & $\bar{\beta}=0.9046$ & &\\
\hline
$\alpha=0.80$ & $\bar{\alpha}=0.8544$ & $\bar{\alpha}=0.9635$ & \multirow{2}{*}{100} & \multirow{2}{*}{100}\\
$\beta=0.80$ & $\bar{\beta}=0.8980$ & $\bar{\beta}=0.9158$ & &\\
\hline
$\alpha=0.85$ & $\bar{\alpha}=0.9011$ & $\bar{\alpha}=0.9726$ & \multirow{2}{*}{96} & \multirow{2}{*}{100}\\
$\beta=0.85$ & $\bar{\beta}=0.9205$ & $\bar{\beta}=0.9253$ & &\\
\hline
$\alpha=0.90$ & $\bar{\alpha}=0.9489$ & $\bar{\alpha}=0.9907$ & \multirow{2}{*}{97} & \multirow{2}{*}{100} \\
$\beta=0.90$ & $\bar{\beta}=0.9436$ & $\bar{\beta}=0.9398$ & &\\
\hline
$\alpha=0.95$ & $\bar{\alpha}=0.9834$ & $\bar{\alpha}=0.9977$ & \multirow{2}{*}{98} & \multirow{2}{*}{100} \\
$\beta=0.95$ & $\bar{\beta}=0.9683$ & $\bar{\beta}=0.9574$ & &\\
\hline
\end{tabular}
\end{table}

\begin{table}
\caption{The quality levels achieved by HYBR on DS and AB.}
\vspace{-0.1cm}
\centering
\label{tab:hyb-quality}
\begin{tabular}{|c|c|c|c|c|}
\hline
Quality & \multicolumn{2}{c|}{Quality Levels of Results} & \multicolumn{2}{c|}{Success rate}\\
\cline{2-5}
Requirement & DS & AB  & \ DS\ \  & AB  \\
\hline
$\alpha=0.70$ & $\bar{\alpha}=0.8649$ & $\bar{\alpha}=0.9304$ & \multirow{2}{*}{100} & \multirow{2}{*}{100} \\
$\beta=0.70$ & $\bar{\beta}=0.8365$ & $\bar{\beta}=0.8306$ & & \\
\hline
$\alpha=0.75$ & $\bar{\alpha}=0.8347$ & $\bar{\alpha}=0.9717$ & \multirow{2}{*}{100} & \multirow{2}{*}{100}\\
$\beta=0.75$ & $\bar{\beta}=0.8573$ & $\bar{\beta}=0.8589$ & &\\
\hline
$\alpha=0.80$ & $\bar{\alpha}=0.8535$ & $\bar{\alpha}=0.9632$ & \multirow{2}{*}{100} & \multirow{2}{*}{100}\\
$\beta=0.80$ & $\bar{\beta}=0.8937$ & $\bar{\beta}=0.8946$ & &\\
\hline
$\alpha=0.85$ & $\bar{\alpha}=0.9015$ & $\bar{\alpha}=0.9898$ & \multirow{2}{*}{95} & \multirow{2}{*}{100}\\
$\beta=0.85$ & $\bar{\beta}=0.9171$ & $\bar{\beta}=0.9160$ & &\\
\hline
$\alpha=0.90$ & $\bar{\alpha}=0.9487$ & $\bar{\alpha}=0.9957$ & \multirow{2}{*}{97} & \multirow{2}{*}{100} \\
$\beta=0.90$ & $\bar{\beta}=0.9425$ & $\bar{\beta}=0.9327$ & &\\
\hline
$\alpha=0.95$ & $\bar{\alpha}=0.9834$ & $\bar{\alpha}=0.9991$ & \multirow{2}{*}{97} & \multirow{2}{*}{100} \\
$\beta=0.95$ & $\bar{\beta}=0.9679$ & $\bar{\beta}=0.9521$ & &\\
\hline
\end{tabular}
\end{table}

\begin{figure}
\setlength{\abovecaptionskip}{\figcaptionspace}
\begin{minipage}[t]{0.5\textwidth}
\centering
\subfigure[Human Cost.]
{\includegraphics[width=0.48\linewidth]{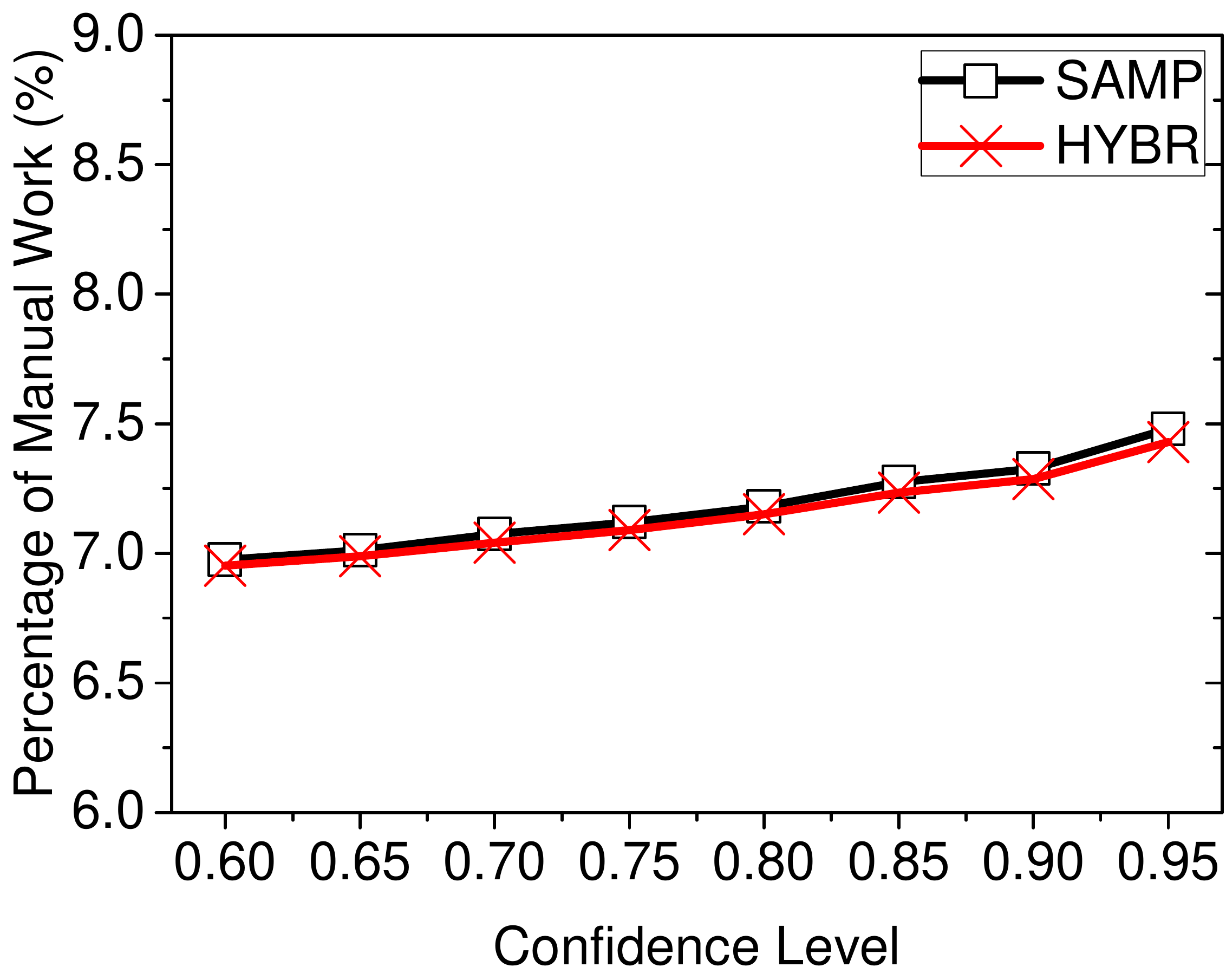}}
\subfigure[Success rate.]
{\includegraphics[width=0.48\linewidth]{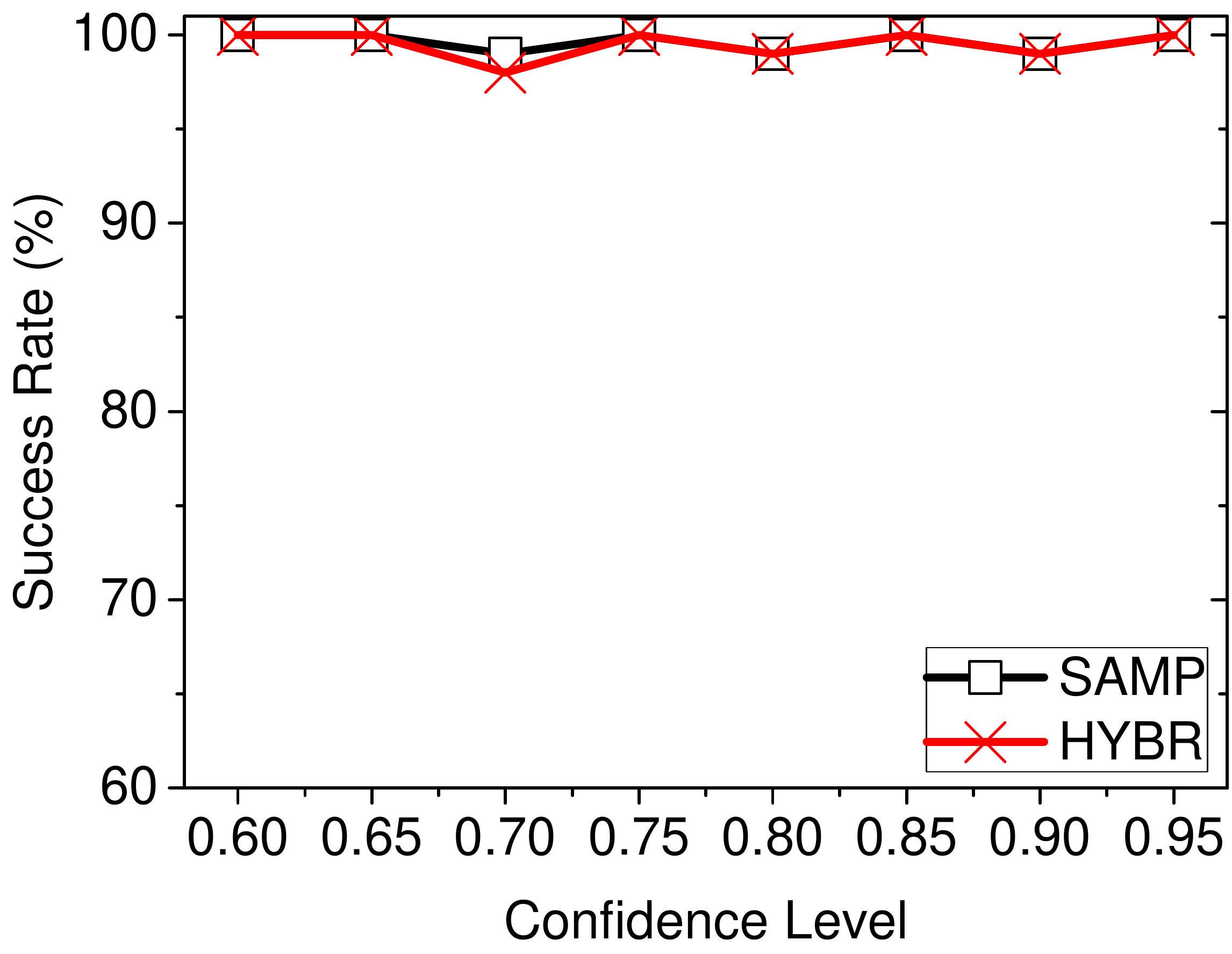}}
\caption{Varying confidence level on DS (with $\alpha=0.9$ and $\beta=0.9$).}
\label{fig:ds-confidence}
\vspace{0.4cm}
\end{minipage}

\begin{minipage}[t]{0.5\textwidth}
\centering
\subfigure[Human Cost.]
{\includegraphics[width=0.48\linewidth]{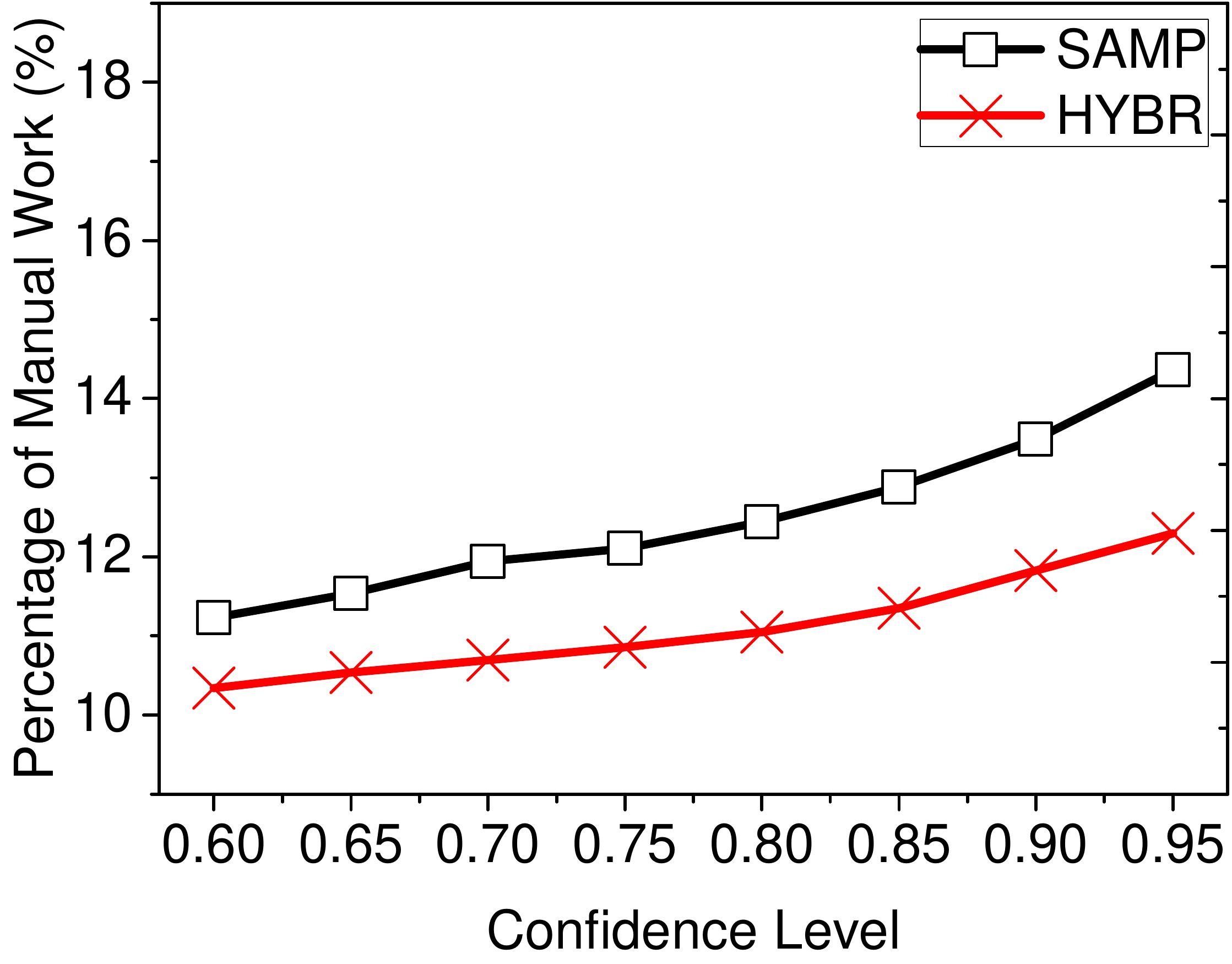}}
\subfigure[Success rate.]
{\includegraphics[width=0.48\linewidth]{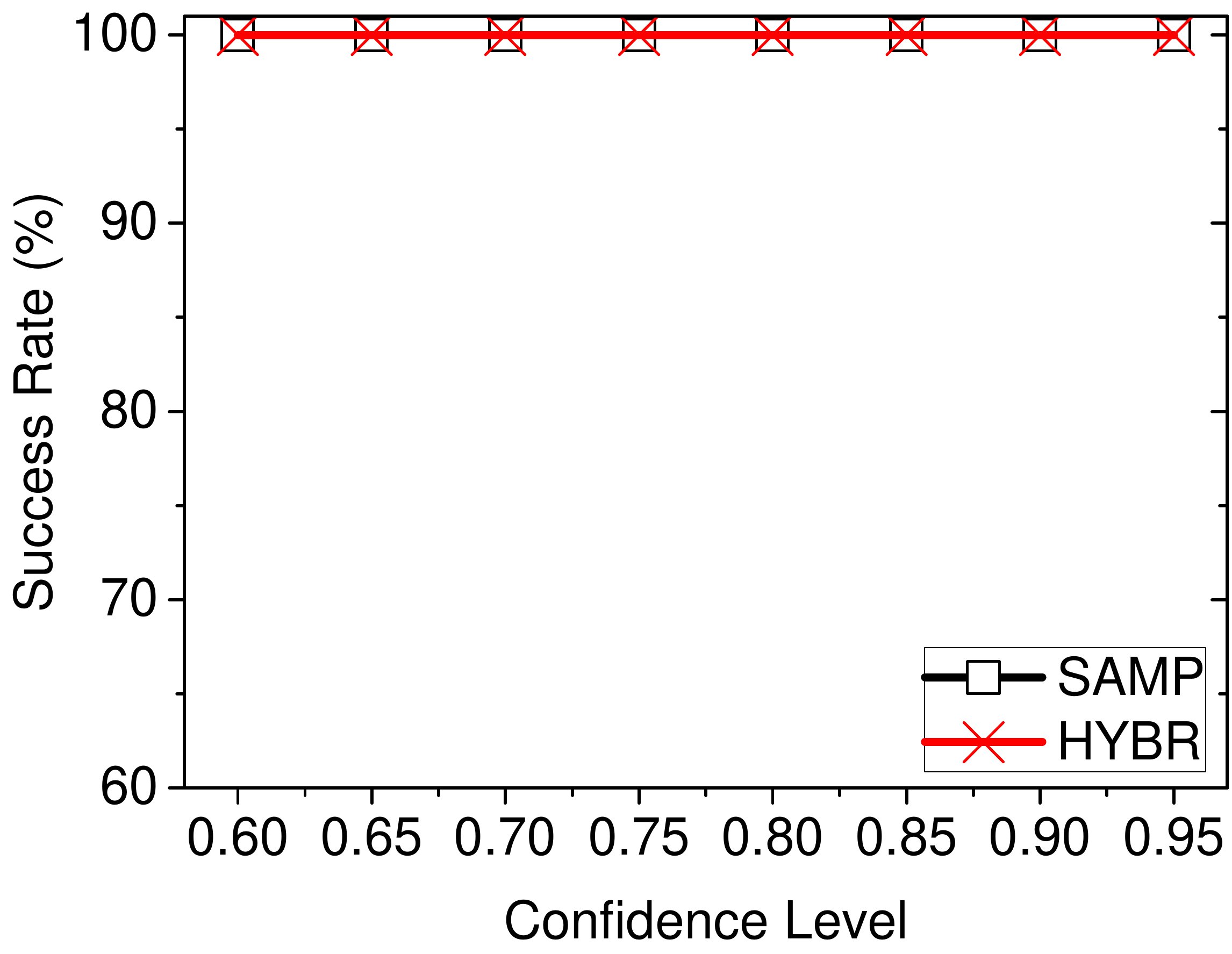}}
\caption{Varying confidence level on AB (with $\alpha=0.9$ and $\beta=0.9$).}
\label{fig:ab-confidence}
\end{minipage}
\vspace{-0.55cm}
\end{figure}

  The comparative results on the two real datasets are presented in Figure~\ref{fig:realdata-evaluation}. On both datasets,  the confidence level for SAMP and HYBR is set to 0.9. Note that for SAMP and HYBR, different runs may generate different solutions due to sampling randomness. Their reported results are therefore the averages over 100 runs. It can be observed that on both datasets, the baseline approach (BASE) requires more human cost than the partial-sampling approach (SAMP). This is mainly due to BASE's conservative estimations of the match proportions of $D_-$ and $D_+$. The more aggressive SAMP approach achieves better performance by effectively reducing their estimation margins. On DS, HYBR performs roughly the same as SAMP; on AB, HYBR however clearly outperforms SAMP. The results on AB show that HYBR can achieve better performance than SAMP by using the better of both BASE and SAMP estimates. It can also be observed that given the same quality requirement, AB requires more human cost than DS. This result should not be surprising given that AB is a more challenging workload than DS. Finally, it is worthy to point out that on both datasets, the required human cost only increases modestly with quality requirement. With both precision and recall guarantees set at 0.9, DS and AB require only around 7\% and 12\% manual work respectively if performed by HYBR.

  We also report the achieved quality levels of different approaches. Note that BASE generates only one HUMO solution on each dataset. Its achieved quality levels on DS and AB are presented in Table~\ref{tab:con-quality}. It can be observed that all the BASE solutions successfully meet the specified quality requirement. Similarly, the achieved quality levels of SAMP and HYBR on DS and AB are presented in Table~\ref{tab:agg-quality} and Table~\ref{tab:hyb-quality} respectively. For SAMP and HYBR, we also report their success rates (to meet quality requirement) of multiple runs besides the averaged precision and recall levels. It can be observed that on both averaged quality and success rate, SAMP and HYBR achieve levels well above what was required in most cases.

  Finally, we evaluate how the required human cost and the success rate of SAMP and HYBR vary with different confidence levels. The required precision and recall levels are both set to 0.9. The detailed results on DS and AB are presented in Figure~\ref{fig:ds-confidence} and Figure~\ref{fig:ab-confidence} respectively. It is clearly observable that the required human cost only increases modestly with the confidence level. SAMP and HYBR's achieved success rates are always above the specified confidence levels. In most cases, the margins between them are considerable. These experimental results demonstrate the robustness of the Gaussian process in approximating match proportions in real application scenarios.

\subsubsection{On Synthetic Datasets}

  Firstly, we fix the parameter value of $\sigma$ at 0.1 and vary the parameter value of $\tau$ from 8 to 18 to test the approaches' performance on the datasets with different match proportion functions. Secondly, we fix the parameter value of $\tau$ at 14 and vary the parameter value of $\sigma$ from 0.1 to 0.5 to test their performance sensitivities to different match proportion irregularities. In both cases, the required precision and recall levels are set to be 0.9. The confidence level of SAMP and HYBR is set at 0.9.

\begin{figure}
\setlength{\abovecaptionskip}{\figcaptionspace}
\centering
\subfigure[Human Cost.]
{\includegraphics[width=0.3\linewidth]{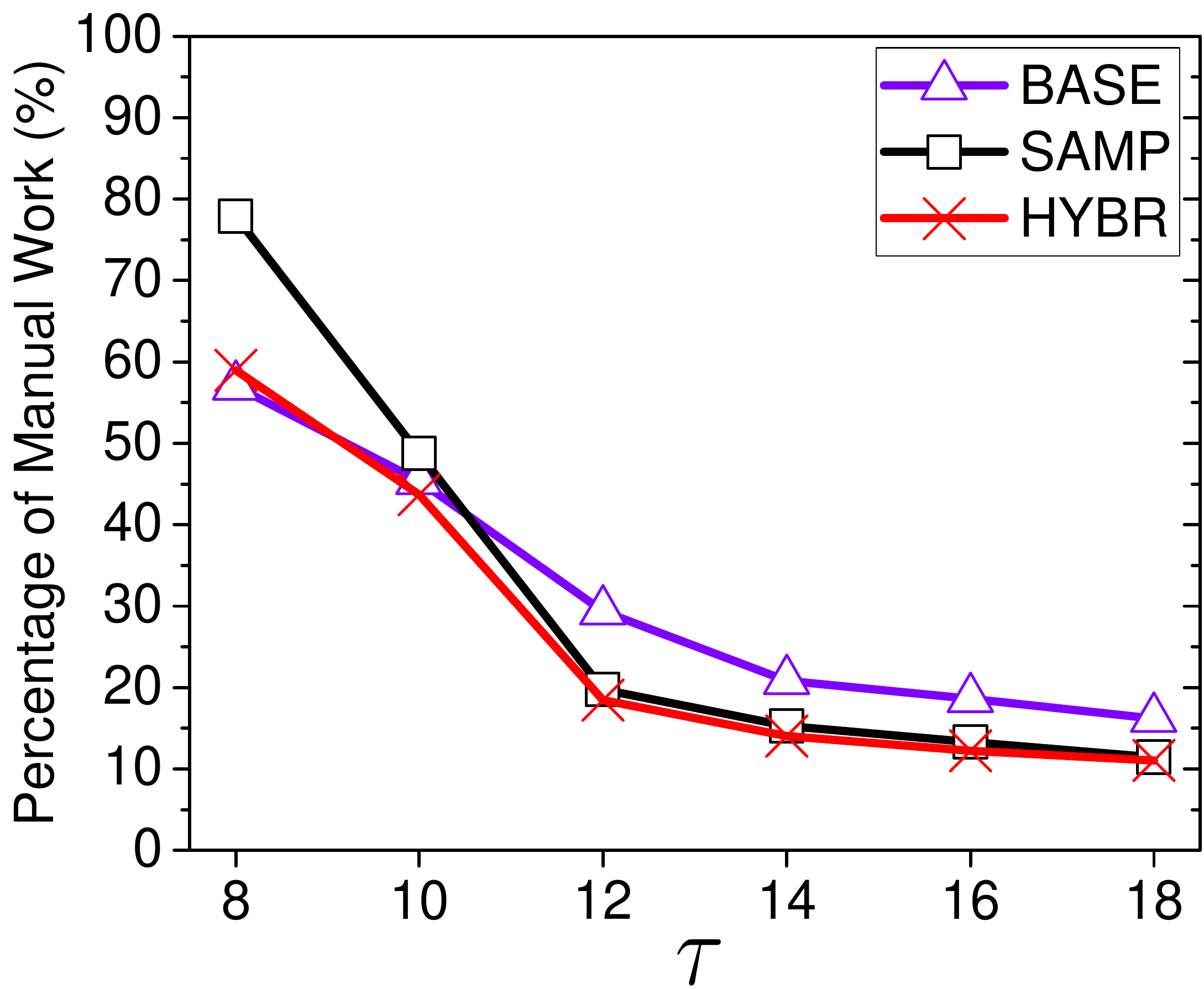}}
\subfigure[Precision level.]
{\label{fig:synthetic-noise01Precision}
\includegraphics[width=0.3\linewidth]{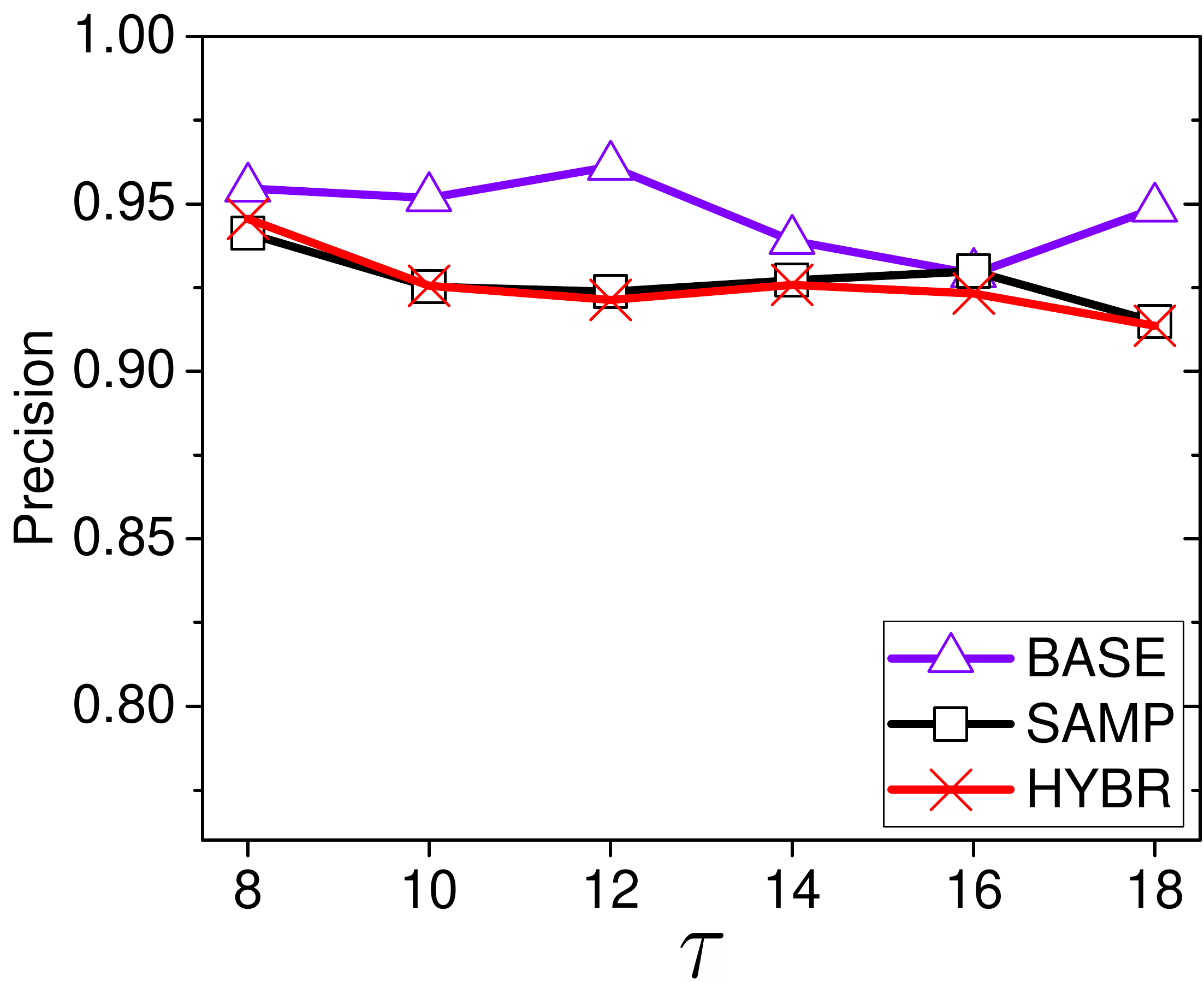}}
\subfigure[Recall level.]
{\label{fig:synthetic-noise01Recall}
\includegraphics[width=0.3\linewidth]{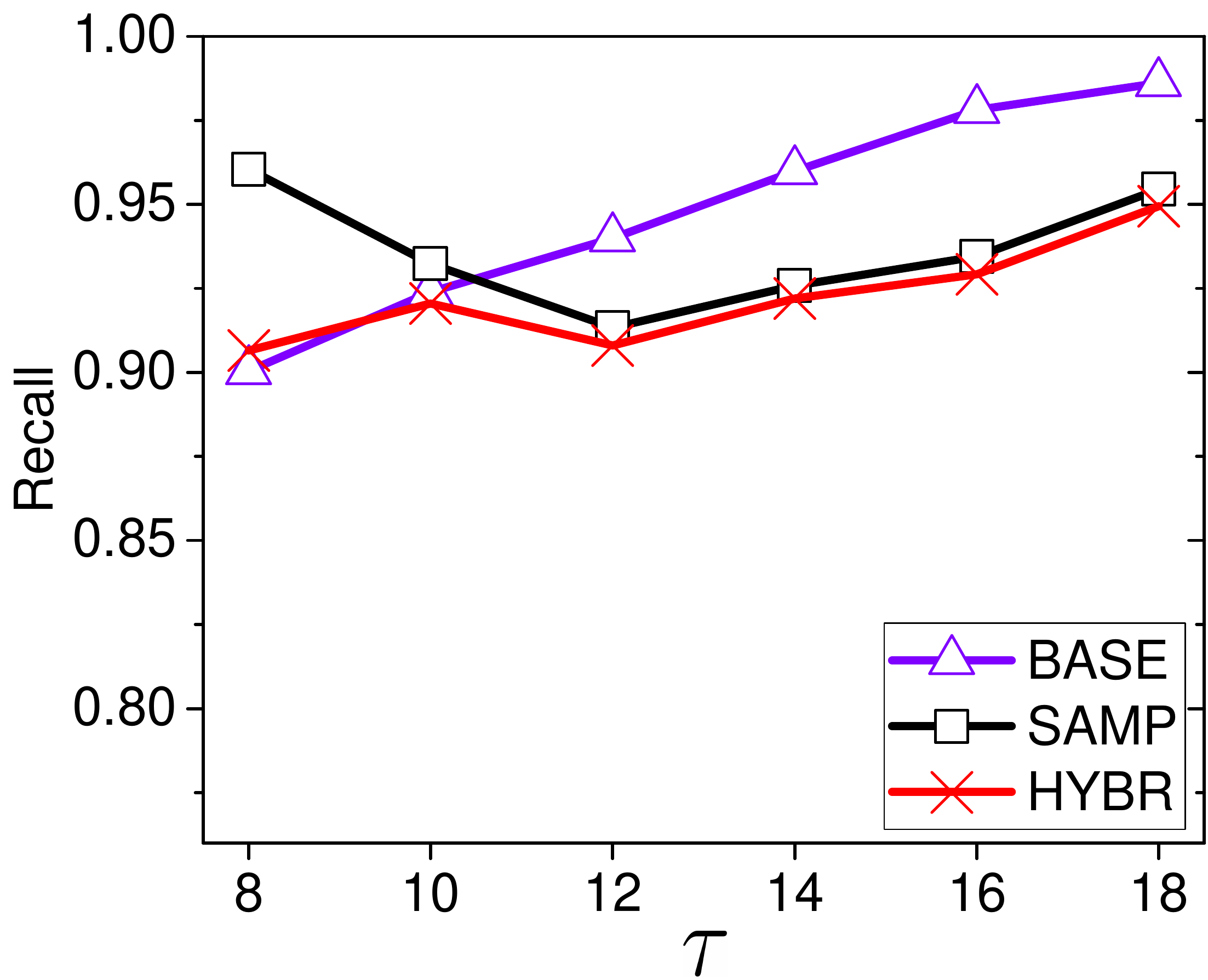}}
\caption{Varying $\tau$ (steepness) of the logistic curve on the synthetic datasets.}
\label{fig:synthetic-experiment}
\vspace{-0.3cm}
\end{figure}

   The detailed evaluation results for the first case are presented in Figure~\ref{fig:synthetic-experiment}. As expected, all the approaches require lesser manual work as $\tau$ is set larger. The results also clearly show that HYBR can effectively use the better of both BASE and SAMP estimates to improve performance. When $\tau\leq10$, BASE requires less manual work than SAMP. When $\tau> 10$, BASE instead requires more manual work than SAMP. However, HYBR can achieve whichever better of BASE and SAMP at all the settings of $\tau$. All the achieved precisions and recalls are observed to be above the required level of 0.9.

\begin{figure}
\setlength{\abovecaptionskip}{\figcaptionspace}
\centering
\subfigure[Human Cost.]
{\includegraphics[width=0.3\linewidth]{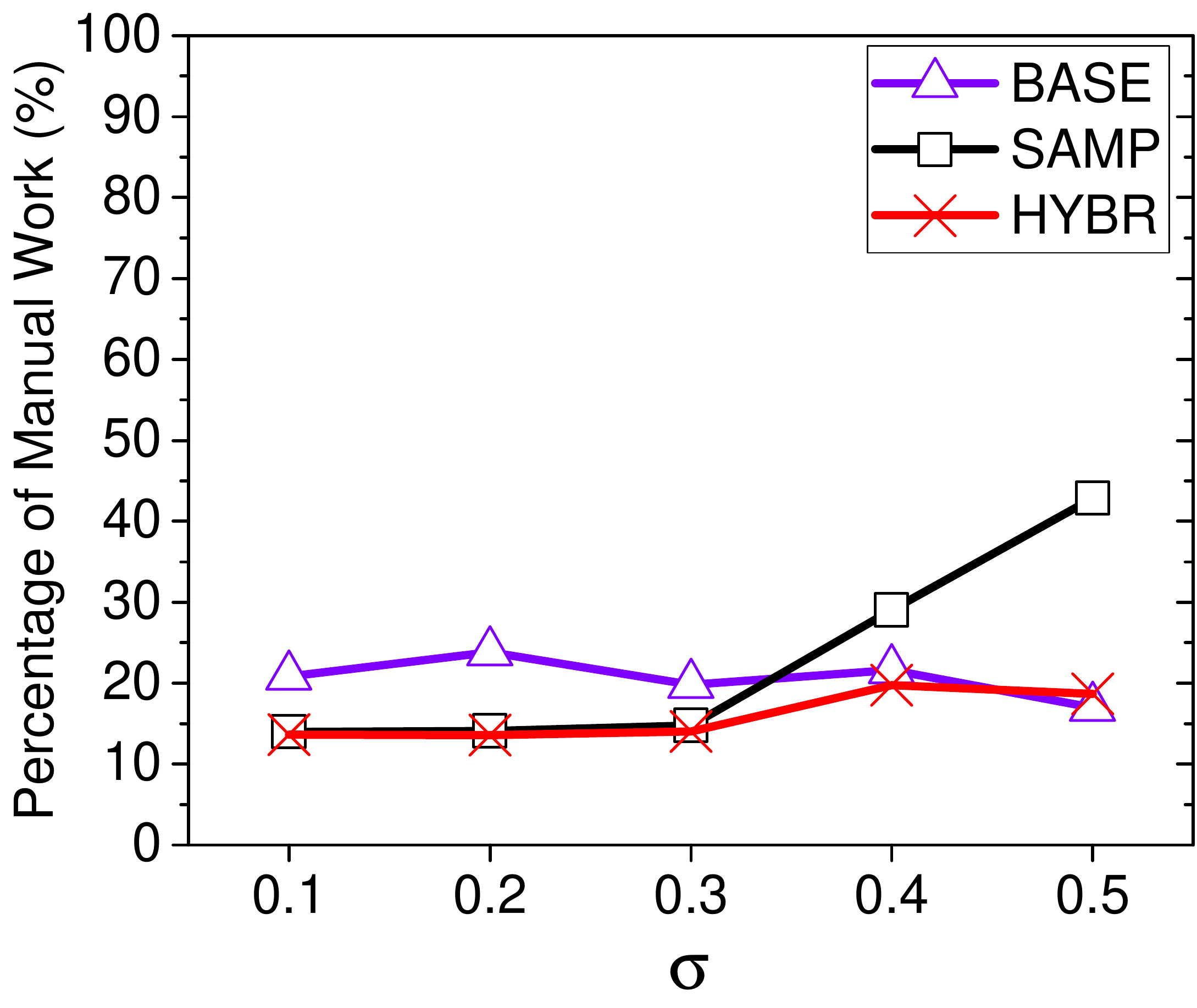}}
\subfigure[Precision level.]
{\includegraphics[width=0.3\linewidth]{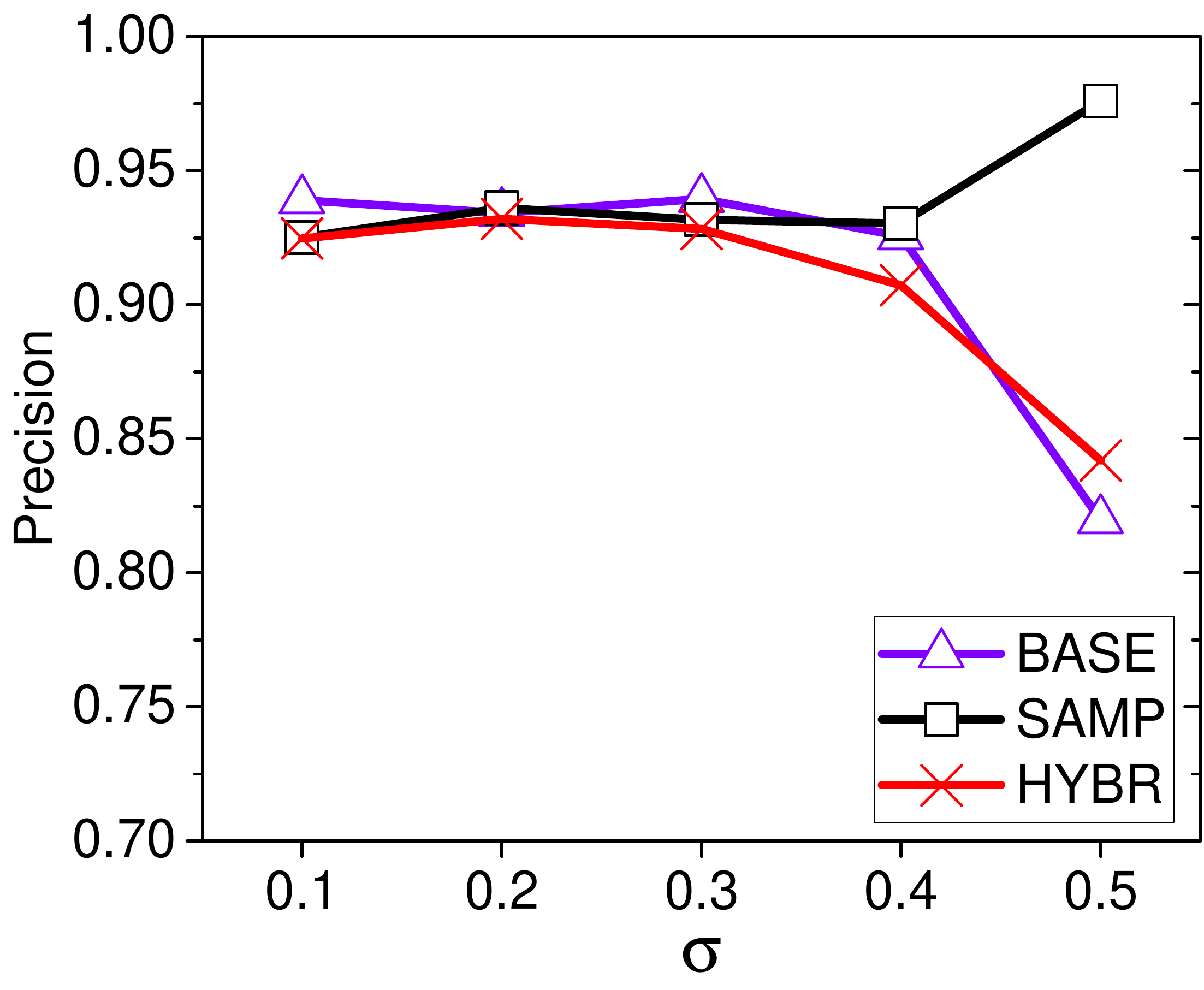}}
\subfigure[Recall level.]
{\includegraphics[width=0.3\linewidth]{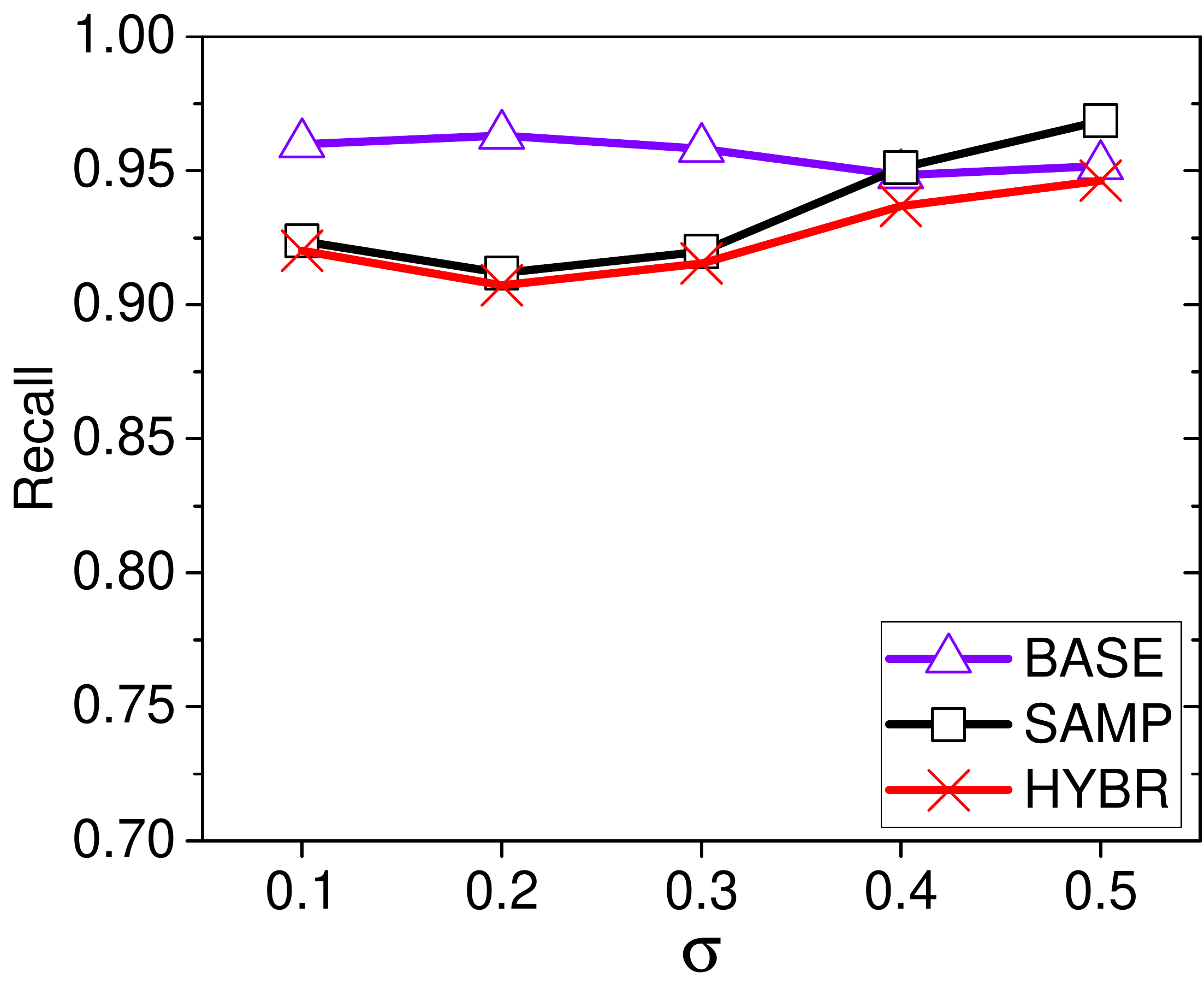}}
\caption{Varying $\sigma$ (variance) of match proportions on the synthetic datasets.}
\label{fig:synthetic-noise-level}
\vspace{-0.55cm}
\end{figure}

 The detailed evaluation results for the second case are presented in Figure~\ref{fig:synthetic-noise-level}. As expected, the required manual workload for SAMP and HYBR generally increases as $\sigma$ is set larger. Similar to what was observed in Figure~\ref{fig:synthetic-experiment}, HYBR achieves the best performance among them by effectively using the better of both BASE and SAMP estimates. With $\sigma\leq 0.4$, all the three approaches can meet the quality requirement. With $\sigma=0.5$, SAMP still manages to meet the quality requirement, but BASE and HYBR fails on precision. This is due to the fact that with $\sigma=0.5$, the monotonicity assumption of precision does not hold true anymore on the synthetic dataset. The effectiveness of SAMP to enforce quality guarantees in the big-variance case of $\sigma=0.5$ also validates the performance resilience of the Gaussian process.

\subsection{Comparison with State-Of-The-Art} \label{sec:other-comparison}

\begin{table}
\begin{minipage}[t]{0.5\textwidth}
\centering
\caption{Performance comparison between HUMO and ACTL on DS. }
\vspace{-0.1cm}
\label{tab:compare-act-ds}
\begin{tabular}{|c|c|c|c|c|c|}
\hline
Target & \multicolumn{2}{c|}{Achieved Recall} & \multicolumn{2}{c|}{$\psi (\%)$} & \multirow{2}{*}{$\frac{\Delta \psi}{100 \cdot \Delta Recall}$}\\
\cline{2-5}
Precision & HUMO & ACTL & HUMO & ACTL & \\
\hline
0.75 & 0.8573 & 0.8210 & 4.94 & 4.08 & 0.2373\\
0.80 & 0.8937 & 0.7953 & 5.52 & 4.10 & 0.1439\\
0.85 & 0.9171 & 0.7786 & 6.20 & 3.73 & 0.1779\\
0.90 & 0.9425 & 0.7124 & 7.34 & 3.63 & 0.1614\\
0.95 & 0.9679 & 0.6502 & 10.05 & 3.01 & 0.2217\\
\hline
\end{tabular}
\vspace{0.5cm}
\end{minipage}
\begin{minipage}[t]{0.5\textwidth}
\centering
\caption{Performance comparison between HUMO and ACTL on AB.}
\vspace{-0.1cm}
\label{tab:compare-act-ab}
\begin{tabular}{|c|c|c|c|c|c|}
\hline
Target & \multicolumn{2}{c|}{Achieved Recall} & \multicolumn{2}{c|}{$\psi (\%)$} & \multirow{2}{*}{$\frac{\Delta \psi}{100 \cdot \Delta Recall}$}\\
\cline{2-5}
Precision & HUMO & ACTL & HUMO & ACTL & \\
\hline
0.75 & 0.8589 & 0.1968 & 6.83 & 0.30 & 0.0985\\
0.80 & 0.8946 & 0.1594 & 7.91 & 0.26 & 0.1040\\
0.85 & 0.9160 & 0.1379 & 9.31 & 0.28 & 0.1161\\
0.90 & 0.9327 & 0.1173 & 11.82 & 0.20 & 0.1426\\
0.95 & 0.9521 & 0.0966 & 16.60 & 0.19 & 0.1918\\
\hline
\end{tabular}
\end{minipage}
\vspace{-0.5cm}
\end{table}

 In this subsection, we compare HUMO with the state-of-the-art alternative (ACTL) based on active learning on the two real datasets. We have implemented both techniques proposed in \cite{arasu2010active} and \cite{bellare2012active} respectively. Our experiments showed that they perform similarly on the achieved quality and required manual work. Their detailed performance comparisons can be found in our technical report \cite{chen2017humoreport}. Here, we present the comparative evaluation results between HUMO and the technique proposed in \cite{arasu2010active}. As \cite{arasu2010active}, we employ Jaccard similarity on attributes used in Subsection~\ref{sec:HUMO-experiment} as the similarity space for ACTL. On DS, the used attributes are {\em title} and {\em authors}; on AB, they are {\em product name} and {\em product description}. ACTL uses sampling to estimate the achieved precision level of a given classification solution; therefore it also requires manual work.

 The performance comparisons between HUMO and ACTL on the DS and AB are presented in Table~\ref{tab:compare-act-ds} and Table~\ref{tab:compare-act-ab} respectively, in which $\psi$ represents the percentage of manual work, and $\Delta$ denotes the performance difference between the two methods on a specified metric. The required precision and recall levels are set to be the same for HUMO. Note that ACTL can not enforce recall level. At each given precision level, we record HUMO and ACTL's differences on achieved recall and consumed human cost. It can be observed that the achieved recall level of ACTL generally decreases with the specified precision level. In all the test cases, HUMO achieves higher recall levels than ACTL. We also record the additional human cost required by HUMO for the absolute recall improvement of $1\%$ over ACTL (at the last columns of Table~\ref{tab:compare-act-ds} and Table~\ref{tab:compare-act-ab}). It can be observed that the cost generally increases with the required precision level. With both precision and recall set at the high level of 0.9, the cost is as low as 0.1614\% on DS and 0.1426\% on AB.

\begin{figure}
\setlength{\abovecaptionskip}{\figcaptionspace}
\centering
\includegraphics[width=0.55\linewidth]{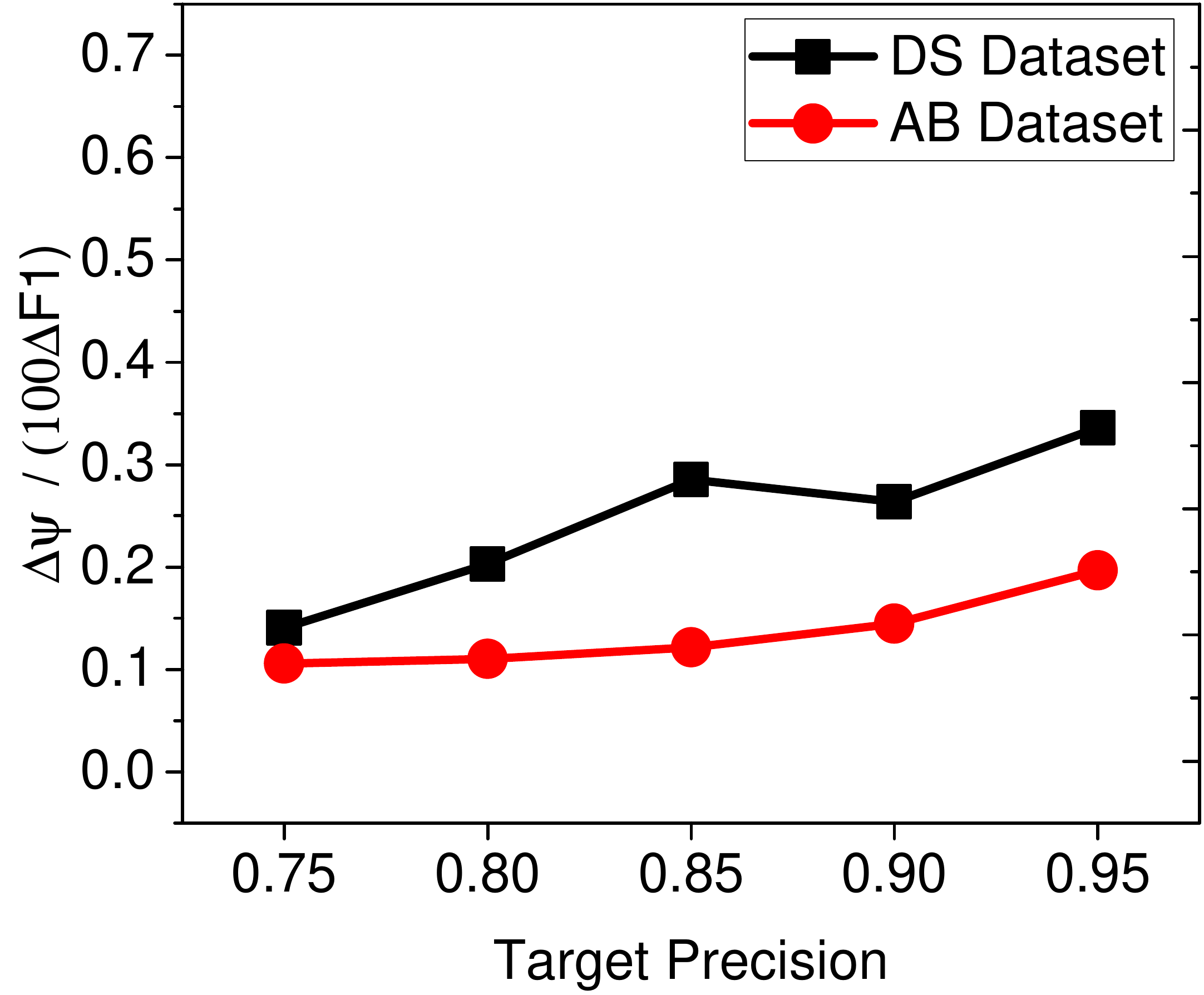}
\caption{The percentage of manual work incurred by HUMO for 1\% absolute improvement in F1 score over ACTL.}
\label{fig:compare-act}
\vspace{-0.2cm}
\end{figure}

  Note that given the same precision requirement, ACTL and HUMO might actually achieve different precisions. Therefore, we also compare their performance on the metric of F1 and record the additional human cost required by HUMO for the absolute F1 improvement of $1\%$ over ACTL. The detailed results on both datasets are presented in Figure~\ref{fig:compare-act}. Similar to what was observed in Table~\ref{tab:compare-act-ds} and Table~\ref{tab:compare-act-ab}, the additional human cost generally increases with the specified precision level. On DS, the additional human cost of HUMO for 1\% increase in F1 score is maximally 0.35\%. On AB, it is as low as 0.21\%.
Along with the results presented in Table~\ref{tab:compare-act-ds} and Table~\ref{tab:compare-act-ab}, these results clearly demonstrate that compared with ACTL, HUMO can effectively improve the resolution quality with reasonable ROI in terms of human cost.

\begin{figure}
\setlength{\abovecaptionskip}{\figcaptionspace}
\centering
{\includegraphics[width=0.55\linewidth]{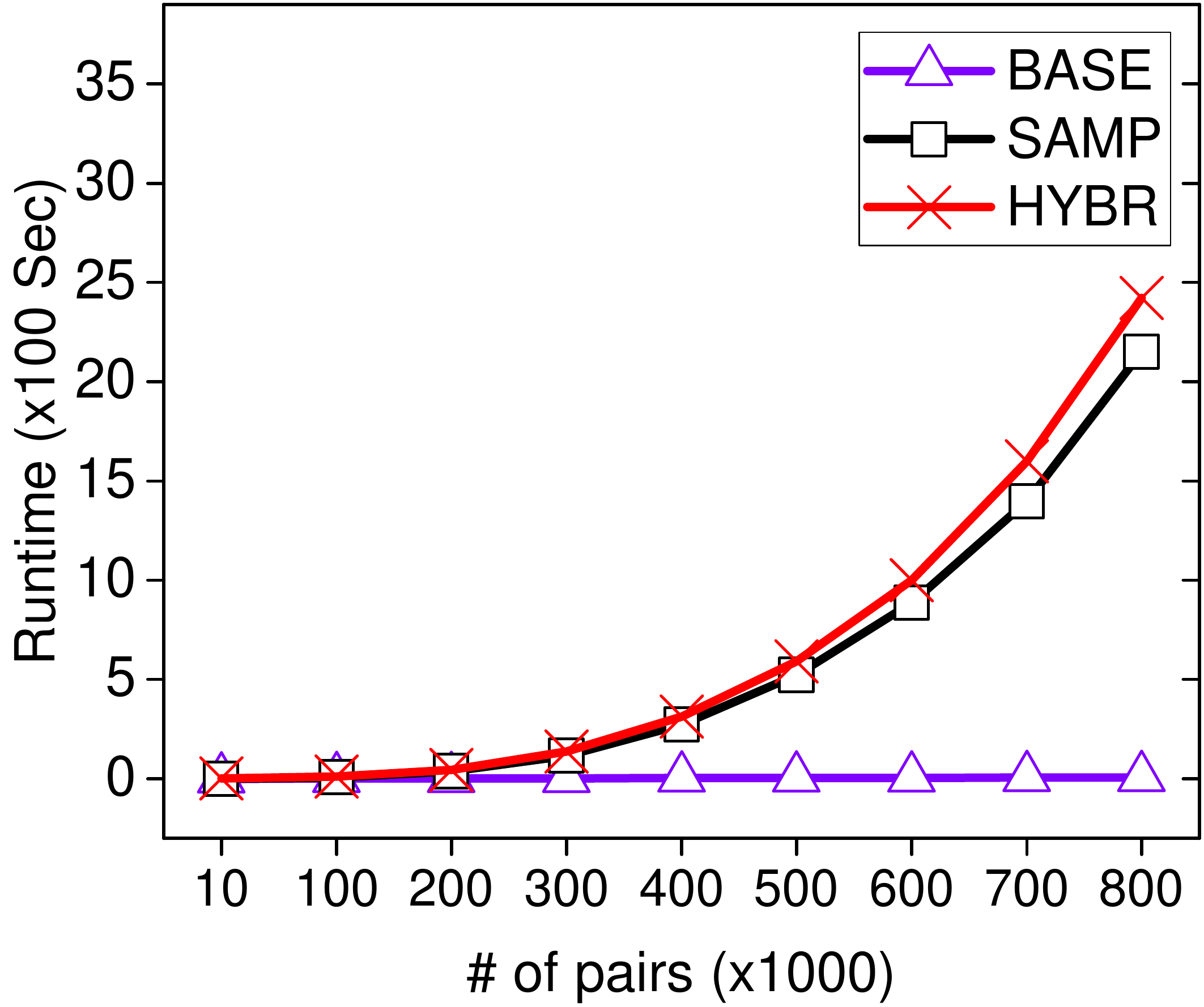}}
\caption{Varying data size on the synthetic datasets.}
\label{fig:scalability}
\vspace{-0.6cm}
\end{figure}

\subsection{Efficiency Evaluation} \label{sec:scalability}

\begin{table}
\vspace{-0.4cm}
\centering
\caption{Efficiency evaluation on DS and AB datasets.}
\vspace{-0.1cm}
\label{tab:efficiency-evaluation}
\begin{tabular}{|c|c|c|c|c|}
\hline
\multirow{2}{*}{Dataset} & \multirow{2}{*}{$\#$ of pairs} & \multicolumn{3}{c|}{Runtime (Sec)} \\
\cline{3-5}
&& BASE & SAMP & HYBR \\
\hline
DS & 100077 & 0.969 & 6.506 & 7.580 \\
\hline
AB & 313040 & 3.062 & 20.921 & 53.503 \\
\hline
\end{tabular}
\end{table}

The machine runtime of HUMO on two real datasets are presented in Table~\ref{tab:efficiency-evaluation}. Note that the reported runtime does not include the time consumed by data preprocessing and the latency incurred by human verification. It can be observed that as indicated by their analyzed complexity, SAMP and HYBR consume considerably more time than BASE. We also evaluate their scalability with varying data size. We generated the test datasets of different sizes using the synthetic data generator. The detailed results are presented in Figure~\ref{fig:scalability}. It can be observed that the runtime of BASE increases only marginally with the increasing dataset size. The runtimes of SAMP and HYBR increase more dramatically but in a polynomial-time manner as dictated by their analyzed complexity. Since efficiency of human work can be expected to be much lower than that of machine computation, machine efficiency of HUMO should not be a concern in practical scenarios.

%% file: 6-conclusion.tex
\section{Conclusion and Future Work} \label{sec:conclusion}
   In this paper, we have proposed a human and machine cooperation framework, HUMO, for entity resolution. It represents a new paradigm that enables a flexible mechanism for comprehensive quality control at both precision and recall levels. Our extensive experiments on real and synthetic datasets have also validated its efficacy.

   Future work can be pursued in two directions. Firstly, for large datasets, crowdsourcing may be the only feasible solution for human workload. It is interesting to integrate HUMO into existing crowdsourcing platforms. On crowdsourcing platforms, monetary cost may be a more appropriate metric of human cost than the number of manually inspected pairs used in this paper. Secondly, as a general paradigm, HUMO can be potentially applied to other challenging classification tasks requiring high quality guarantees (e.g. financial fraud detection \cite{ngai2011application} and malware detection \cite{ye2017survey}). It is interesting to investigate its efficacy on them in future work.

%% file: 7-acknowledgements.tex
\section*{Acknowledgment}
\vspace{-1pt}
This work was supported by the Ministry of Science and Technology of China, National Key Research and Development Program (2016YFB1000703), NSF of China (61732014, 61332006, 61472321, 61502390 and 61672432).

%% file: ms.bbl
\begin{thebibliography}{10}
\providecommand{\url}[1]{#1}
\csname url@samestyle\endcsname
\providecommand{\newblock}{\relax}
\providecommand{\bibinfo}[2]{#2}
\providecommand{\BIBentrySTDinterwordspacing}{\spaceskip=0pt\relax}
\providecommand{\BIBentryALTinterwordstretchfactor}{4}
\providecommand{\BIBentryALTinterwordspacing}{\spaceskip=\fontdimen2\font plus
\BIBentryALTinterwordstretchfactor\fontdimen3\font minus
  \fontdimen4\font\relax}
\providecommand{\BIBforeignlanguage}[2]{{%
\expandafter\ifx\csname l@#1\endcsname\relax
\typeout{** WARNING: IEEEtran.bst: No hyphenation pattern has been}%
\typeout{** loaded for the language `#1'. Using the pattern for}%
\typeout{** the default language instead.}%
\else
\language=\csname l@#1\endcsname
\fi
#2}}
\providecommand{\BIBdecl}{\relax}
\BIBdecl

\bibitem{christen2012data}
P.~Christen, \emph{Data matching: concepts and techniques for record linkage,
  entity resolution, and duplicate detection}.\hskip 1em plus 0.5em minus
  0.4em\relax Springer Science \& Business Media, 2012.

\bibitem{fan2009reasoning}
W.~Fan, X.~Jia, J.~Li, and S.~Ma, ``Reasoning about record matching rules,''
  \emph{Proceedings of the VLDB Endowment}, vol.~2, no.~1, pp. 407--418, 2009.

\bibitem{li2015rule}
L.~Li, J.~Li, and H.~Gao, ``Rule-based method for entity resolution,''
  \emph{IEEE Transactions on Knowledge and Data Engineering}, vol.~27, no.~1,
  pp. 250--263, 2015.

\bibitem{singh2017generating}
R.~Singh, V.~Meduri, A.~Elmagarmid, S.~Madden, P.~Papotti, J.-A.
  Quian{\'e}-Ruiz, A.~Solar-Lezama, and N.~Tang, ``Generating concise entity
  matching rules,'' \emph{Proceedings of the ACM International Conference on
  Management of Data (SIGMOD)}, pp. 1635--1638, 2017.

\bibitem{fellegi1969theory}
I.~P. Fellegi and A.~B. Sunter, ``A theory for record linkage,'' \emph{Journal
  of the American Statistical Association}, vol.~64, no. 328, pp. 1183--1210,
  1969.

\bibitem{sarawagi2002interactive}
S.~Sarawagi and A.~Bhamidipaty, ``Interactive deduplication using active
  learning,'' \emph{Proceedings of the 8th ACM International Conference on
  Knowledge Discovery and Data Mining (SIGKDD)}, pp. 269--278, 2002.

\bibitem{kouki2017collective}
P.~Kouki, J.~Pujara, C.~Marcum, L.~Koehly, and L.~Getoor, ``Collective entity
  resolution in familial networks,'' \emph{IEEE International Conference on
  Data Mining (ICDM)}, pp. 227--236, 2017.

\bibitem{arasu2010active}
A.~Arasu, M.~G{\"o}tz, and R.~Kaushik, ``On active learning of record matching
  packages,'' \emph{Proceedings of the ACM International Conference on
  Management of Data (SIGMOD)}, pp. 783--794, 2010.

\bibitem{bellare2012active}
K.~Bellare, S.~Iyengar, A.~G. Parameswaran, and V.~Rastogi, ``Active sampling
  for entity matching,'' \emph{Proceedings of the 18th ACM international
  Conference on Knowledge Discovery and Data Mining (SIGKDD)}, pp. 1131--1139,
  2012.

\bibitem{li2016crowdsourced}
G.~Li, J.~Wang, Y.~Zheng, and M.~J. Franklin, ``Crowdsourced data management: A
  survey,'' \emph{IEEE Transactions on Knowledge and Data Engineering},
  vol.~28, no.~9, pp. 2296--2319, 2016.

\bibitem{wang2012crowder}
J.~Wang, T.~Kraska, M.~J. Franklin, and J.~Feng, ``Crowder: Crowdsourcing
  entity resolution,'' \emph{Proceedings of the VLDB Endowment}, vol.~5,
  no.~11, pp. 1483--1494, 2012.

\bibitem{whang2013question}
S.~E. Whang, P.~Lofgren, and H.~Garcia-Molina, ``Question selection for crowd
  entity resolution,'' \emph{Proceedings of the VLDB Endowment}, vol.~6, no.~6,
  pp. 349--360, 2013.

\bibitem{vesdapunt2014crowdsourcing}
N.~Vesdapunt, K.~Bellare, and N.~Dalvi, ``Crowdsourcing algorithms for entity
  resolution,'' \emph{Proceedings of the VLDB Endowment}, vol.~7, no.~12, pp.
  1071--1082, 2014.

\bibitem{gokhale2014corleone}
C.~Gokhale, S.~Das, A.~Doan, J.~F. Naughton, N.~Rampalli, J.~Shavlik, and
  X.~Zhu, ``Corleone: Hands-off crowdsourcing for entity matching,''
  \emph{Proceedings of the ACM International Conference on Management of Data
  (SIGMOD)}, pp. 601--612, 2014.

\bibitem{mozafari2014scaling}
B.~Mozafari, P.~Sarkar, M.~Franklin, M.~Jordan, and S.~Madden, ``Scaling up
  crowd-sourcing to very large datasets: a case for active learning,''
  \emph{Proceedings of the VLDB Endowment}, vol.~8, no.~2, pp. 125--136, 2014.

\bibitem{wang2015crowd}
S.~Wang, X.~Xiao, and C.-H. Lee, ``Crowd-based deduplication: An adaptive
  approach,'' \emph{Proceedings of the ACM International Conference on
  Management of Data (SIGMOD)}, pp. 1263--1277, 2015.

\bibitem{chai2016cost}
C.~Chai, G.~Li, J.~Li, D.~Deng, and J.~Feng, ``Cost-effective crowdsourced
  entity resolution: A partial-order approach,'' \emph{Proceedings of the ACM
  International Conference on Management of Data (SIGMOD)}, pp. 969--984, 2016.

\bibitem{chen2017humo}
Z.~Chen, Q.~Chen, and Z.~Li, ``A human-and-machine cooperative framework for
  entity resolution with quality guarantees,'' \emph{IEEE 33rd International
  Conference on Data Engineering (ICDE), Demo paper}, pp. 1405--1406, 2017.

\bibitem{elmagarmid2007duplicate}
A.~K. Elmagarmid, P.~G. Ipeirotis, and V.~S. Verykios, ``Duplicate record
  detection: A survey,'' \emph{IEEE Transactions on Knowledge and Data
  Engineering}, vol.~19, no.~1, pp. 1--16, 2007.

\bibitem{christophides2015entity}
V.~Christophides, V.~Efthymiou, and K.~Stefanidis, ``Entity resolution in the
  web of data,'' \emph{Synthesis Lectures on the Semantic Web}, vol.~5, no.~3,
  pp. 1--122, 2015.

\bibitem{singla2006entity}
P.~Singla and P.~Domingos, ``Entity resolution with markov logic,'' \emph{IEEE
  6th International Conference on Data Mining (ICDM)}, pp. 572--582, 2006.

\bibitem{whang2013pay}
S.~E. Whang, D.~Marmaros, and H.~Garcia-Molina, ``Pay-as-you-go entity
  resolution,'' \emph{IEEE Transactions on Knowledge and Data Engineering},
  vol.~25, no.~5, pp. 1111--1124, 2013.

\bibitem{altowim2014progressive}
Y.~Altowim, D.~V. Kalashnikov, and S.~Mehrotra, ``Progressive approach to
  relational entity resolution,'' \emph{Proceedings of the VLDB Endowment},
  vol.~7, no.~11, pp. 999--1010, 2014.

\bibitem{lacoste2013sigma}
S.~Lacoste-Julien, K.~Palla, A.~Davies, G.~Kasneci, T.~Graepel, and
  Z.~Ghahramani, ``Sigma: Simple greedy matching for aligning large knowledge
  bases,'' pp. 572--580, 2013.

\bibitem{verroios2017waldo}
V.~Verroios, H.~Garcia-Molina, and Y.~Papakonstantinou, ``Waldo: An adaptive
  human interface for crowd entity resolution,'' pp. 1133--1148, 2017.

\bibitem{kopcke2010evaluation}
H.~K{\"o}pcke, A.~Thor, and E.~Rahm, ``Evaluation of entity resolution
  approaches on real-world match problems,'' \emph{Proceedings of the VLDB
  Endowment}, vol.~3, no. 1-2, pp. 484--493, 2010.

\bibitem{chen2017humoreport}
\BIBentryALTinterwordspacing
Z.~Chen, Q.~Chen, F.~Fan, Y.~Wang, Z.~Wang, Y.~Nafa, Z.~Li, H.~Liu, and W.~Pan,
  ``Enabling quality control for entity resolution: A human and machine
  cooperation framework (technical report),'' Tech. Rep., 2017. [Online].
  Available: \url{http://www.wowbigdata.com.cn/HUMO/technical-report.pdf}
\BIBentrySTDinterwordspacing

\bibitem{cochran1977sampling}
W.~G. Cochran, \emph{Sampling techniques}, 3rd~ed.\hskip 1em plus 0.5em minus
  0.4em\relax John Wiley \& Sons, 1977.

\bibitem{rasmussen2006gaussian}
C.~E. Rasmussen and C.~K. Williams, \emph{Gaussian processes for machine
  learning}.\hskip 1em plus 0.5em minus 0.4em\relax MIT press Cambridge, 2006.

\bibitem{ngai2011application}
E.~Ngai, Y.~Hu, Y.~Wong, Y.~Chen, and X.~Sun, ``The application of data mining
  techniques in financial fraud detection: A classification framework and an
  academic review of literature,'' \emph{Decision Support Systems}, vol.~50,
  no.~3, pp. 559--569, 2011.

\bibitem{ye2017survey}
Y.~Ye, T.~Li, D.~Adjeroh, and S.~S. Iyengar, ``A survey on malware detection
  using data mining techniques,'' \emph{ACM Computing Surveys (CSUR)}, vol.~50,
  no.~3, p.~41, 2017.

\end{thebibliography}
